\documentclass[aps,pra,reprint,superscriptaddress]{revtex4-1}
\usepackage{graphicx}
\usepackage{bbm}

\begin{document}

\title{Optimal sequential measurements for bi-partite state discrimination}
\author{Sarah Croke}
\author{Stephen M. Barnett}
\author{Graeme Weir}
\email{g.weir.2@research.gla.ac.uk}
\affiliation{School of Physics and Astronomy, University of Glasgow, Glasgow G12 8QQ, UK}


\date{\today}

\newcommand{\bra}[1]{\langle #1|}
\newcommand{\ket}[1]{|#1\rangle}
\newcommand{\braket}[2]{\langle #1|#2\rangle}

\begin{abstract}
State discrimination is a useful test problem with which to clarify the power and limitations of different classes of measurement. We consider the problem of discriminating between given states of a bi-partite quantum system via sequential measurement of the subsystems, with classical feed-forward of measurement results. Our aim is to understand when sequential measurements, which are relatively easy to implement experimentally, perform as well, or almost as well as optimal joint measurements, which are in general more technologically challenging. We construct conditions that the optimal sequential measurement must satisfy, analogous to the well-known Helstrom conditions for minimum error discrimination in the unrestricted case. We give several examples and compare the optimal probability of correctly identifying the state via  global versus sequential measurement strategies. 
\end{abstract}

\maketitle

\section{Introduction}
The problem of quantum state discrimination is most naturally thought of as a task in quantum communications, although it also has applications elsewhere in quantum information theory and quantum metrology \cite{Barnett09review,Bae15review,Chefles00review,Bergou10review,SteveBook}. The communications scenario is as follows: a sender, traditionally called Alice, chooses a quantum state $\rho_i$ drawn from a given set $\{ \rho_j \}$ with associated \emph{a priori} probabilities $\{ p_j \}$, and sends a system prepared in this state to a receiver, Bob. Bob knows the allowed set of states and their associated probabilities, and his task is to determine which state was sent, thereby recovering the message sent by Alice. The task was first considered in the pioneering work of Helstrom, Holevo, and others in the late 60s and 70s \cite{Helstrom67,Helstrom68,Holevo73,Yuen75,Davies78,Helstrom76,Holevo82}. Various strategies exist, each optimising a different figure of merit (see e.g. \cite{Davies78,Levitin95,Sasaki99,Ivanovic87,Dieks88,Peres88,Jaeger95,Chefles98,MaxConf}), and for arguably the simplest such figure, minimising the probability of error in identifying the state, necessary and sufficient conditions for a quantum measurement strategy to be optimal are known \cite{Holevo73,Yuen75}.

More recently, state discrimination has proved a useful test problem with which to clarify the power and limitations of different classes of measurement. For information encoded across multiple quantum systems, the ability to measure jointly is strictly more powerful (but in general technologically more challenging) than the ability to measure each subsystem independently, even if many rounds of classical communication between systems are allowed. Intuitively, one might expect the difference in performance to be more pronounced when information is encoded in entangled states. That this is not necessarily the case was first revealed through two state discrimination problems. The first, so-called ``non-locality without entanglement", gave a set of multi-partite orthogonal \emph{product} states between which perfect discrimination is not possible using only local measurements and classical communication \cite{Bennett99}. The second, complementary and no less surprising, showed that any two orthogonal pure states, regardless of entanglement or multi-partite structure, may be perfectly discriminated using only sequential measurement, i.e. local measurement on each system, with classical feed-forward \cite{Walgate00}. This was later extended to show that any two in general non-orthogonal pure states may be discriminated optimally by sequential measurement of the subsystems, according to the commonly used minimum error \cite{Virmani01} and unambiguous discrimination strategies \cite{Chen01,Chen02,Ji05}.

Beyond the two state examples, the situation becomes much less clear: for the next simplest example of discriminating three possible qubit states given two copies, it was postulated by Peres and Wootters in 1991 that local measurement was strictly weaker than joint measurement on both copies \cite{Peres91}, and only twenty years later was it finally proved that such a gap exists for this problem, for the minimum error strategy \cite{Chitambar13}.

In this paper we consider sequential measurements on a bi-partite system; i.e. subsystem A and B are measured in turn, and the choice of measurement performed on subsystem B is allowed to depend in general on the result of measurement of A. This is often a physically relevant class of measurement; for example if A and B are in different labs it is easy to imagine that feedforward of measurement results from lab A to lab B would be practical but many rounds of classical communication could become unfeasible. Alternatively if A and B interact only weakly or not at all (e.g. photons), joint measurements are difficult to perform, while classical feed-forward from one detector to another apparatus is relatively easily achieved with current technology (see e.g. \cite{Lu10} for such an experiment in the state discrimination context). It is natural then to ask how well information can be retrieved with this restriction on the measurement strategy that may be employed. Further, implementations of joint measurement strategies for extracting information may provide applications for small quantum processors \cite{BlumeKohout13}, and it is useful to understand when the additional experimental challenge of joint measurement may provide a significant advantage over local measurement strategies. For simplicity, we restrict to bipartite instead of the more general multipartite state discrimination.

We begin with the case where the bipartite state is simply a two-copy state. We construct necessary conditions that a given sequential measurement must satisfy to be optimal in the sense of minimising the error in determining the state, analogous to the well-known Helstrom conditions \cite{Holevo73,Yuen75}. We further find a condition which is both necessary and sufficient, but which requires optimisation over an arbitrary measurement on one subsystem. We illustrate the two-copy case through the example of the trine states considered in \cite{Peres91,Chitambar13}, and give the probabilities of correctly identifying the state for sequential and global strategies, as well as discussing features of the optimal measurements in each case.

We extend the discussion to arbitrary bi-partite states, and as an example give the optimal sequential strategies for discriminating three Bell states, and for discriminating the so-called domino states introduced by Bennett \emph{et al.} in \cite{Bennett99}. Finally we discuss an interpretation for our necessary and sufficient condition in terms of a related discrimination problem.

\section{Review: Helstrom conditions \label{helstrom}}
We first recall the minimum error problem, where there are no restrictions on the allowed measurement: a quantum system is prepared in one of a known set of states $\{ \rho_i \}$ with associated probabilities $\{ p_i \}$. Any physically allowed measurement may be represented by a POVM (positive operator-valued measure) \cite{Peres95}, also referred to as a POM (probability operator measure) \cite{Helstrom76}, that is, a set of Hermitian operators $\{ \pi_i \}$ satisfying:
\begin{eqnarray*}
\pi_i &\geq& 0, \\
\sum_i \pi_i &=& \mathbbmss{1}.
\end{eqnarray*}
For a measurement described by operators $\{ \pi_i \}$, if outcome $i$ is taken to indicate state $\rho_i$, the probability of correctly identifying the state is given by:
\begin{equation}
P_{\rm corr} = \sum_i p_i {\rm Tr}(\rho_i \pi_i).
\end{equation}
The operators $\{ \pi_i \}$ describing the optimal measurement satisfy the following conditions \cite{SteveBook,Holevo73,Yuen75,Barnett09}:
\begin{eqnarray}
\sum_i p_i \rho_i \pi_i - p_j \rho_j &\geq& 0, \quad \forall j \label{MEcond} \\
\pi_i (p_i \rho_i - p_j \rho_j) \pi_j &=& 0, \quad \forall i,j. \label{MEnecc}
\end{eqnarray}
It is worth noting that the conditions are not independent, as the second follows from the first. Condition (\ref{MEnecc}) may be thought of as analogous to the condition in an optimisation problem that the first derivative vanish at a stationary point, while condition (\ref{MEcond}) is analogous to the second derivative condition: it is the sign of the second derivative which determines whether the corresponding point is a local maximum or local minumum. Condition (\ref{MEnecc}) is therefore necessary but not sufficient for $\{ \pi_i \}$ to be an optimal measurement, however (\ref{MEcond}) is both necessary and sufficient. We give here a sketch of the proof, following the treatment of \cite{Barnett09}, which is extended to the sequential case in the rest of the paper.

If $\{ \pi_i \}$ is optimal then for all other physically allowed measurements $\{ \pi_i^\prime \}$ we require
$$
P_{\rm corr} (\{ \pi_i \}) \geq P_{\rm corr} (\{ \pi_i^\prime \}).
$$
From this we obtain
\begin{eqnarray}
\sum_i p_i {\rm Tr}(\rho_i \pi_i) - \sum_j p_j {\rm Tr}(\rho_j \pi_j^\prime) & \geq & 0 \nonumber \\
\sum_j {\rm Tr} \left[ \left(\sum_i p_i \rho_i \pi_i - p_j \rho_j \right) \pi_j^\prime \right] & \geq & 0. \label{sufficient}
\end{eqnarray}
Note that for positive operators $A$, $B$ it is always true that ${\rm Tr}(AB) \geq 0$, which may be seen by evaluating the trace in the eigenbasis of $A$:
\begin{equation}
{\rm Tr} \left( AB \right) = \sum_i \bra{a_i} A B \ket{a_i} = \sum_i a_i \bra{a_i} B \ket{a_i} \geq 0,
\label{AB}
\end{equation}
where $a_i \geq 0$ are the eigenvalues of $A$, $\{ \ket{a_i} \}$ are the eigenkets of $A$, and the inequality follows from the positivity of $B$. As $\pi_j^\prime$ is a positive operator it is therefore clear that condition (\ref{MEcond}) is sufficient in order for the inequality (\ref{sufficient}) to be satisfied. That this condition is also necessary may be shown by introducing the Hermitian operators
$$
G_j = \sum_i p_i \frac{1}{2} \left\{ \rho_i, \pi_i \right\} - p_j \rho_j.
$$
Now if $\exists \ket{\lambda}$ such that $\bra{\lambda} G_j \ket{\lambda} < 0$, the variation
\begin{equation}
\pi_i^\prime = (\mathbbmss{1} - \epsilon \ket{\lambda} \bra{\lambda}) \pi_i (\mathbbmss{1} - \epsilon \ket{\lambda} \bra{\lambda}) + \epsilon (2 + \epsilon) \ket{\lambda} \bra{\lambda} \delta_{ij}
\label{variation}
\end{equation}
results in a measurement with higher probability of success than $\{ \pi_i \}$, which therefore cannot be an optimal measurement. Finally it is possible to show \cite{Barnett09} that
$$
\sum_j G_j \pi_j = 0
$$
and thus
$$
\sum_i p_i \frac{1}{2} \left\{ \rho_i, \pi_i \right\} = \sum_j p_j \rho_j \pi_j.
$$
Thus the requirement $G_j \geq 0$ reduces to condition (\ref{MEcond}), which is therefore both necessary and sufficient.

It is useful to denote $\Gamma = \sum_i p_i \rho_i \pi_i$. We finish by noting that for an optimal measurement $\{ \pi_j^\prime \}$, we require
\[
P_{\rm corr} = {\rm Tr}(\Gamma) = \sum_j p_j {\rm Tr}(\rho_j \pi_j^\prime),
\]
and therefore
\[
\sum_j {\rm Tr}((\Gamma - p_j \rho_j ) \pi_j^\prime) = 0.
\]
As discussed above, for positive operators $A$, $B$, ${\rm Tr}(AB) \geq 0$, and it is clear from eqn (\ref{AB}) that equality holds if and only if $AB = 0$. Thus we require that each term in the sum be identically zero, which further requires
\begin{equation}
(\Gamma - p_j \rho_j ) \pi_j^\prime = 0, \quad \forall j \label{altcond}
\end{equation}
for any optimal measurement $\{ \pi_j^\prime \}$. This is an alternative necessary (but not sufficient) condition, and is sometimes useful for finding optimal measurements. It also implies, on summing over $j$, that $\Gamma$ is unique, $\Gamma = \sum_j p_j \rho_j \pi_j^\prime$ for any optimal $\{ \pi_j^\prime \}$ (see also \cite{Bae13}).

\section{Two copy state discrimination with sequential measurement \label{twocopy}}
\subsection{Necessary conditions}
\begin{figure}[h!]
\begin{center}
\includegraphics[width=80mm]{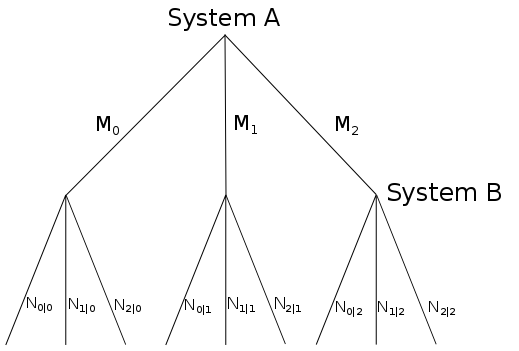}
\end{center}
\caption{Probability tree showing sequential measurement notation: the measurement described by POVM $\{M_j\}$ is performed on system $A$. Given outcome $j$, the measurement described by POVM $\{ N_{i|j} \}$ is performed on system $B$.}
\label{Tree}
\end{figure}

Now let us consider the two copy case, with sequential measurement. Suppose therefore we are provided with two copies of a state drawn from a known set $\{ \rho_i \}$ with associated probabilities $\{ p_i \}$. The allowed measurement procedures are as follows: make a measurement described by some POVM $\{ M_j^A \}$ on system $A$; given outcome $j$ make a measurement on system $B$, as shown in the tree in Figure \ref{Tree}.
As the choice of measurement on system $B$ can in general depend on the outcome of measurement on $A$, we denote the associated POVM $\{ N_{i|j}^B \}$, where for all $i$ and $j$, $N_{i|j}^B \geq 0$ and for each $j$
$$
\sum_i N_{i|j}^B = \mathbbmss{1}^B.
$$
The measurement on the joint $AB$ system is thus of the form $\{ \pi_i = \sum_j M_j^A \otimes N_{i|j}^B \}$, with the probability of correctly identifying the state given by:
\begin{eqnarray}
{\rm P}_{\rm corr} &=& \sum_{ij} p_i {\rm Tr}_{AB} \left( \rho_i^A \otimes \rho_i^B M_j^A \otimes N_{i|j}^B \right) \nonumber \\
&=&  \sum_{ij} p_i {\rm Tr}_{A} \left( \rho_i^A M_j^A \right) {\rm Tr}_{B} \left( \rho_i^B N_{i|j}^B \right). \label{Pcorrseq}
\end{eqnarray}
In the following we drop the superscripts $A$, $B$, whenever it is not confusing to do so. We begin by pointing out that each of $\{ M_j \}$, $\{ N_{i|j} \}$ may be interpreted as an optimal measurement for an appropriately defined discrimination problem, as follows. We first note that, given measurement result $j$ on system $A$, we can update the probabilities as follows, using Bayes' rule:
\begin{equation}
{\rm P}(i|M_j) = \frac{{\rm P}(i,M_j)}{{\rm P}(M_j)} = \frac{p_i {\rm Tr}_A(\rho_i M_j)}{\sum_k p_k {\rm Tr}_A(\rho_k M_j)} = p_{i|j}. \label{posteriori}
\end{equation}
Thus given result $j$ on system $A$, the possible states $\{ \rho_i \}$ of system $B$ occur with probabilities $p_{i|j}$. Clearly $\{ N_{i|j} \}$ should thus be optimal for discriminating the states $ \rho_i $ with the updated priors $p_{i|j}$, and thus a necessary condition is
$$
\sum_i p_{i|j} \rho_i N_{i|j} - p_{k|j} \rho_k \geq 0, \quad \forall k,
$$
or equivalently, using eqn. (\ref{posteriori}):
\begin{equation}
\sum_i p_i {\rm Tr}_A(\rho_i M_j) \rho_i N_{i|j} - p_k {\rm Tr}_A(\rho_k M_j) \rho_k \geq 0, \quad \forall k,
\label{Nopt}
\end{equation}
which must hold for each $j$. This set of conditions is necessary, but not sufficient (we haven't done any optimisation over $M_j$). Finally, summing over $j$ gives:
\begin{equation}
{\rm Tr}_A \left( \sum_{i,j} p_i (\rho_i \otimes \rho_i) (M_j \otimes N_{i|j}) - p_k \rho_k \otimes \rho_k \right) \geq 0, \quad \forall k,
\label{Noptsum}
\end{equation}
which is rather similar to the Helstrom condition (\ref{MEcond}), but with a partial trace over system $A$.

Conversely, we can re-write eqn (\ref{Pcorrseq}) as follows:
\begin{eqnarray*}
{\rm P}_{\rm corr} &=& \sum_{j}  {\rm Tr}_{A} \left( \sum_i p_i {\rm Tr}_{B} \left( \rho_i^B N_{i|j}^B \right) \rho_i^A M_j^A \right) \\
 &=& \sum_{j}  c_j {\rm Tr}_{A} \left( \sigma_j^A M_j^A \right),
\end{eqnarray*}
where we have defined:
\begin{eqnarray}
\sigma_j^A &=& \frac{\sum_i p_i {\rm Tr}_{B} \left( \rho_i^B N_{i|j}^B \right) \rho_i^A}{\sum_k p_k {\rm Tr}_{B} \left( \rho_k^B N_{k|j}^B \right)}, \label{sigma} \\
c_j &=& \sum_i p_i {\rm Tr}_{B} \left( \rho_i^B N_{i|j}^B \right). \label{cj}
\end{eqnarray}
We can interpret the trace one operators $\{ \sigma_j \}$ as density operators, and if we further define probabilities $q_j=c_j/\left(\sum_i c_i \right)$, it follows that $\{ M_j^A \}$ must be optimal for discriminating the states $\{ \sigma_j^A \}$ with probabilities $\{ q_j \}$. The Helstrom condition (\ref{MEcond}) then gives:
$$
 \sum_{j}  q_j \sigma_j^A M_j^A - q_k \sigma_k^A \geq 0,
$$
which may be re-written as:
\begin{eqnarray}
 \sum_{j} \left(\sum_i p_i {\rm Tr}_{B} \left( \rho_i^B N_{i|j}^B \right) \rho_i^A \right) M_j^A \nonumber \\
-  \sum_i p_i {\rm Tr}_{B} \left( \rho_i^B N_{i|k}^B \right) \rho_i^A &\geq& 0.
\end{eqnarray}
Finally we obtain
\begin{eqnarray}
{\rm Tr}_B \left( \sum_{ij} p_i (\rho_i^A \otimes \rho_i^B) (M_j^A \otimes N_{i|j}^B) \right. \nonumber \\
 - \left. \sum_i p_i (\rho_i^A \otimes \rho_i^B) (\mathbbmss{1}^A \otimes N_{i|k}^B) \right) &\geq& 0. \label{Mopt}
\end{eqnarray}
Again, this is necessary, but not sufficient (this time we haven't done any optimisation over $N_{i|j}$). One might hope that the conditions (\ref{Nopt},\ref{Mopt}) when taken together are also sufficient, and could then imagine that it may be possible to construct an iterative procedure for numerical solution of the optimization problem. However, this turns out not to be the case; we will return to this point later. Each of conditions (\ref{Nopt},\ref{Mopt}) however have a clear interpretation; note that it might have been expected that $\{ N_{i|j}^B \}$ should be optimal for the updated priors given measurement of $A$; that $M_j^A$ plays a complementary role for a different discrimination problem is less obvious a priori. We return to give an interpretation of this discrimination problem later.

\subsection{A necessary and sufficient condition}
We now turn to the problem of simultaneously optimising both the measurement on $A$ and that on system $B$. We find that the condition
\begin{equation}
\sum_{i,j} p_i {\rm Tr}_B (\rho_i N_{i|j}) \rho_i M_j - \sum_{k} p_k {\rm Tr}_B (\rho_k \widetilde{N}_k) \rho_k \geq 0
\label{seqcond}
\end{equation}
where $\{ \widetilde{N}_k \}$ is any physically allowed measurement on system $B$ is both necessary and sufficient for optimality of $\{ \pi_i = \sum_j M_j \otimes N_{i|j} \}$. Unfortunately this still contains an arbitrary measurement on system $B$, and thus is not as readily applicable as the original Helstrom conditions to verify optimality of a candidate measurement. Nevertheless we will give examples in which it can be used to prove optimality analytically. We also note that the inclusion of an arbitrary measurement on one subsystem means that analysis beyond the bipartite case becomes complicated and our method is not readily extended to multipartite discrimination.

We begin by proving sufficiency of condition (\ref{seqcond}). If $\{ \pi_i = \sum_j M_j \otimes N_{i|j} \}$ is optimal among sequential measurements, we require
\begin{eqnarray*}
{\rm Tr}_{AB} \left( \sum_{i,j} p_i (\rho_i \otimes \rho_i) (M_j \otimes N_{i|j}) \right) \geq \\  {\rm Tr}_{AB} \left( \sum_{k,l} p_k (\rho_k \otimes \rho_k) (M_l^\prime \otimes N_{k|l}^\prime) \right),
\end{eqnarray*}
for all $\{ \pi_k^\prime = \sum_l M_l^\prime \otimes N_{k|l}^\prime \}$. Inserting the identity $\sum_l M_l^\prime \otimes \mathbbmss{1}$ and re-arranging gives
\begin{eqnarray*}
\sum_l {\rm Tr}_{AB} \left[ \left( \sum_{i,j} p_i (\rho_i \otimes \rho_i) (M_j \otimes N_{i|j}) \right. \right. \\ - \left. \left. \sum_{k} p_k (\rho_k \otimes \rho_k) (\mathbbmss{1} \otimes N_{k|l}^\prime) \right) M_l^\prime \right] &\geq& 0, \\
\sum_l {\rm Tr}_{A} \left[ \left( \sum_{i,j} p_i {\rm Tr}_B (\rho_i N_{i|j}) \rho_i M_j \right. \right. \\ - \left. \left. \sum_{k} p_k {\rm Tr}_B (\rho_k N_{k|l}^\prime) \rho_k \right) M_l^\prime \right] &\geq& 0.
\end{eqnarray*}
Condition (\ref{seqcond}) is therefore sufficient, if $\{ \widetilde{N}_k \}$ is any allowed measurement on $B$.

That condition (\ref{seqcond}) is also necessary may be seen as follows: as in the unrestricted case, we introduce the manifestly Hermitian operator:
$$
\Gamma_{\rm sym}^A = \sum_{i,j} p_i {\rm Tr}_B \left( \rho_i N_{i|j} \right) \frac{1}{2} \{ \rho_i, M_j \}.
$$
Suppose now that there exists some $\ket{\lambda}$ and some $\{ \widetilde{N}_k \}$ such that 
$$
\bra{\lambda} \Gamma_{\rm sym}^A - \sum_{k} p_k {\rm Tr}_B (\rho_k \widetilde{N}_k) \rho_k \ket{\lambda} < 0.
$$
We can construct a variation of $\{ \pi_i = \sum_j M_j \otimes N_{i|j} \}$ as follows:
\begin{eqnarray*}
M_j^\prime &=& (\mathbbmss{1}-\epsilon \ket{\lambda} \bra{\lambda}) M_j (\mathbbmss{1} - \epsilon \ket{\lambda} \bra{\lambda}), \quad 0 \leq j < n \\
N_{i|j}^\prime &=& N_{i|j}, \quad 0 \leq j < n \\
M_{n} &=& \epsilon (2 + \epsilon) \ket{\lambda} \bra{\lambda}, \\
N_{i|n} &=& \widetilde{N}_i,
\end{eqnarray*}
where $0 < \epsilon \ll 1$. Note that if $\{ M_j^A \}$ has $n$ outcomes, the primed measurement on system $A$ has $n+1$ outcomes. Now note that
\begin{eqnarray*}
&& {\rm P_{corr}} \left( \{ M_j^\prime \otimes N_{i|j}^\prime \} \right) = {\rm P_{corr}} \left( \{ M_j \otimes N_{i|j} \} \right) \\
&& - \epsilon {\rm Tr}_{AB}\left( \sum_{i,j} p_i \rho_i \otimes \rho_i \left( \ket{\lambda} \bra{\lambda} M_j + M_j \ket{\lambda} \bra{\lambda} \right) \otimes N_{i|j} \right) \\
&& + 2 \epsilon {\rm Tr}_{AB} \left( \sum_i p_i \rho_i \otimes \rho_i ( \ket{\lambda} \bra{\lambda} \otimes \widetilde{N}_i) \right) + O(\epsilon^2) \\
&& = {\rm P_{corr}} \left( \{ M_j \otimes N_{i|j} \} \right) \\
&& - 2 \epsilon \bra{\lambda} \Gamma_{\rm sym}^A - \sum_i p_i {\rm Tr}_B (\rho_i \widetilde{N}_i) \rho_i \ket{\lambda} + O(\epsilon^2) \\
&& > {\rm P_{corr}} \left( \{ M_j \otimes N_{i|j} \} \right).
\end{eqnarray*}
Finally we note that, by virtue of the fact that $\{ M_j \}$ is an optimal measurement for discriminating the states $\sigma_j$, it follows that $\Gamma_{\rm sym}^A = \Gamma^A$, where $\Gamma^A$ is defined as:
$$
\Gamma^A = \sum_{i,j} p_i {\rm Tr}_B \left( \rho_i N_{i|j} \right) \rho_i M_j.
$$ 
Thus we require
$$
\Gamma^A - \sum_k p_k {\rm Tr}_B \left( \rho_k \widetilde{N}_k \right) \rho_k \geq 0,
$$
which completes our proof.

\section{Example: The double trine ensemble}
As an example we consider the so-called double trine ensemble: two copies of the trine states, for which $\rho_j = \ket{\psi_j} \bra{\psi_j}$, and
\[
\ket{\psi_j} = \frac{1}{\sqrt2} \left( \ket{0} + e^{2 \pi j i/3} \ket{1} \right).
\]
These each occur with prior probabilities $p_j = \frac{1}{3}$, and have the symmetry property
\[
\ket{\psi_{j}} = U^j \ket{\psi_0}
\]
where $U$ is a rotation of $\frac{2 \pi}{3}$ around the $z$-axis in the Bloch sphere.

\subsection{Optimal sequential measurement}
For the two-copy case, Chitambar and Hsieh \cite{Chitambar13} showed that the optimal sequential measurement rules out one state of the three in the first step, and corresponds to the Helstrom measurement to distinguish between the remaining two states in the second step. We first briefly present this optimal measurement, and then use it to demonstrate our conditions.

\begin{figure}[h!]
\begin{center}
\includegraphics[width=80mm]{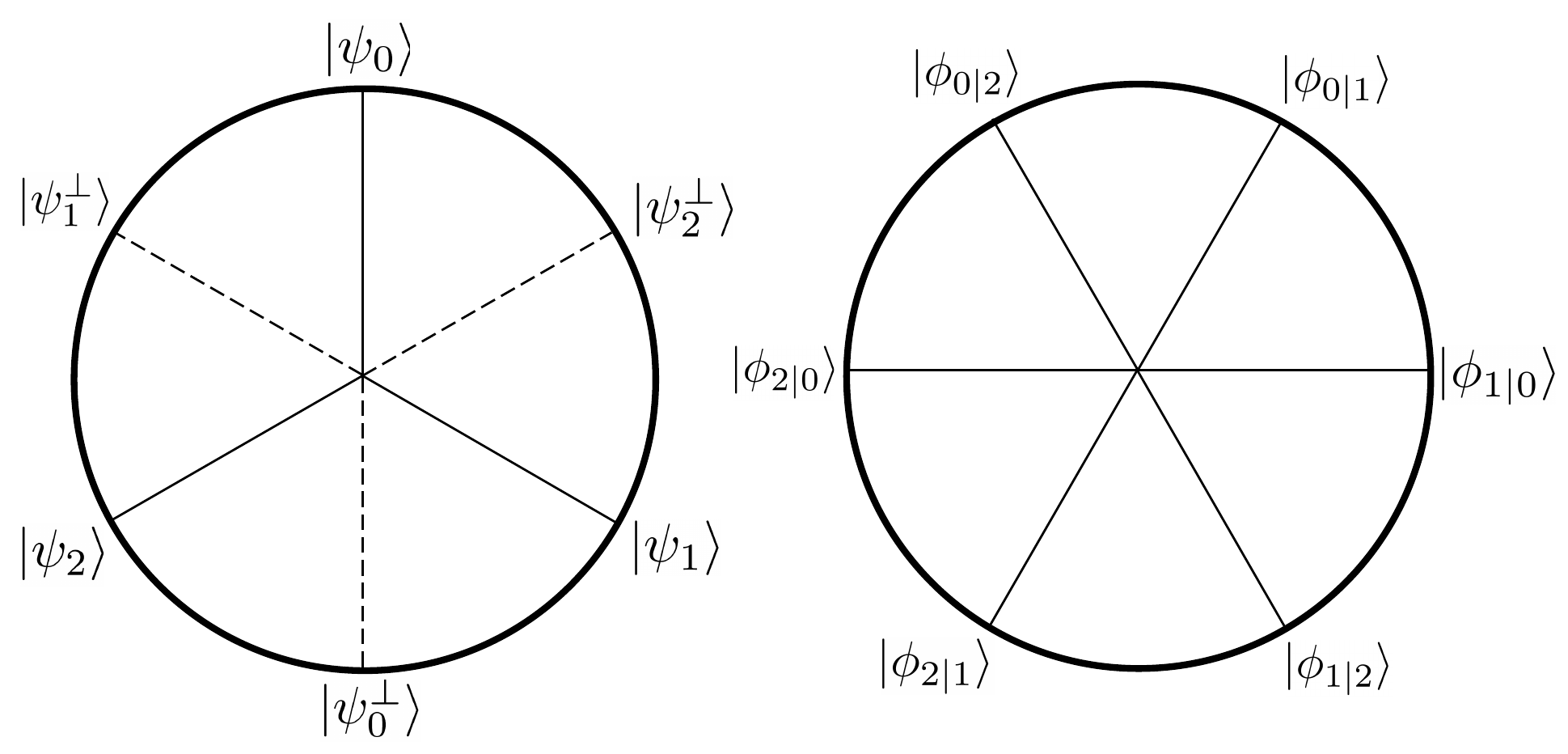}
\end{center}
\caption{Trine and anti-trine states shown in the equator of the Bloch sphere (left). Bases defined by the optimal Helstrom measurements in step two of the optimal sequential measurement procedure for discriminating the two-copy trine ensemble (right).}
\label{BlochFigure}
\end{figure}

The optimal sequential measurement thus makes the measurement $\{ M_j = \frac{2}{3} \ket{\psi_j^\perp} \bra{\psi_j^\perp} \}$ on the first copy, where the states $\{ \ket{\psi_j^\perp} \}$ form the so-called anti-trine ensemble:
\[
\ket{\psi_j^\perp} = \frac{1}{\sqrt2} \left( \ket{0} - e^{2 \pi j i/3} \ket{1} \right).
\]
Following this measurement, the updated priors become $p_{i|j} = \frac{1}{2} (1- \delta_{ij})$, and $\{ N_{i|j} \}$ is then the optimal measurement to distinguish the two remaining equiprobable pure states $\{ \ket{\psi_i}, \ket{\psi_k}, i \neq j \neq k \}$. This is a case of the well-known Helstrom measurement, and is a projective measurement in a basis located symmetrically around the signal states (see e.g. \cite{Barnett09review}). Thus for $i=j$, $N_{i|j} = 0$, and for $i \neq j$ we denote $N_{i|j} = \ket{\phi_{i|j}} \bra{\phi_{i|j}}$, where
\begin{eqnarray*}
\ket{\phi_{1|0}} &=& \frac{1}{\sqrt2} \left( \ket{0} + i \ket{1} \right), \\
\ket{\phi_{2|0}} &=& \frac{1}{\sqrt2} \left( \ket{0} - i \ket{1} \right), \\
\ket{\phi_{0|1}} &=& \frac{1}{\sqrt2} \left( \ket{0} + e^{i \pi/6} \ket{1} \right) = U \ket{\phi_{2|0}}, \\
\ket{\phi_{2|1}} &=& \frac{1}{\sqrt2} \left( \ket{0} - e^{i \pi/6} \ket{1} \right) = U \ket{\phi_{1|0}}, \\
\ket{\phi_{0|2}} &=& \frac{1}{\sqrt2} \left( \ket{0} + e^{-i \pi/6} \ket{1} \right) = U^2 \ket{\phi_{1|0}}, \\
\ket{\phi_{1|2}} &=& \frac{1}{\sqrt2} \left( \ket{0} - e^{-i \pi/6} \ket{1} \right) = U^2 \ket{\phi_{2|0}}.
\end{eqnarray*}
These states, along with the trine and anti-trine states are shown in the Bloch sphere picture in Figure \ref{BlochFigure}.

\subsection{Necessary and sufficient conditions}
We now use this strategy to illustrate the conditions presented in the previous section. From the symmetry we find that ${\rm Tr}(\rho_i N_{i|j}) = p_H (1-\delta_{ij})$, for all $i,j$, where $p_H$ is the probability of success of the Helstrom measurement distinguishing between two equiprobable states with overlap $| \braket{\psi_i}{\psi_k} | =| \braket{\psi_0}{\psi_1} | = 1/2$, i.e. from \cite{Helstrom76}:
$$
p_H = \frac{1}{2} \left(1 + \sqrt{1 - |\braket{\psi_0}{\psi_1}|^2} \right) = \frac{1}{2} \left(1 + \frac{\sqrt{3}}{2} \right).
$$
By construction, this measurement strategy satisfies condition (\ref{Nopt}). To evaluate (\ref{Mopt}) and the necessary and sufficient condition (\ref{seqcond}), we first calculate $\Gamma^A$:
\begin{eqnarray*}
\Gamma^A &=& \sum_{i,j} p_i {\rm Tr} \left( \rho_i N_{i|j} \right) \rho_i M_j \\
&=& \sum_{i,j} \frac{1}{3} p_H (1-\delta_{ij}) \left( \ket{\psi_i} \bra{\psi_i} \right) \left( \frac{2}{3} \ket{\psi_j^\perp} \bra{\psi_j^\perp} \right) \\
&=& \frac{1}{3} p_H \left( \sum_i \ket{\psi_i} \bra{\psi_i} \right) \left(\sum_j \frac{2}{3} \ket{\psi_j^\perp} \bra{\psi_j^\perp} \right) \\
&=& \frac{1}{2} p_H \mathbbmss{1}= \frac{1}{4} \left(1+ \frac{\sqrt{3}}{2} \right) \mathbbmss{1},
\end{eqnarray*}
where in the last line we have used $\sum_j \frac{2}{3} \ket{\psi_j^\perp} \bra{\psi_j^\perp} = \sum_j \frac{2}{3} \ket{\psi_j} \bra{\psi_j} = \mathbbmss{1}$. We first show that the strategy satisfies condition (\ref{Mopt}). We obtain
\begin{eqnarray*}
\sum_i p_i {\rm Tr} \left( \rho_i N_{i|j} \right) \rho_i^A &=& \frac{1}{3} \sum_i p_H (1-\delta_{ij}) \rho_i \\
&=& \frac{1}{2} p_H \left( \mathbbmss{1} -\frac{2}{3} \rho_j \right) 
\end{eqnarray*}
from which it is clear that condition (\ref{Mopt}) is satisfied for each $j$. Finally, to prove that this is indeed the optimal strategy, we must show that it satisfies the necessary and sufficient condition (\ref{seqcond}). As we have shown that $\Gamma^A$ is proportional to the identity, this amounts to showing that for any allowed measurement $\{ \widetilde{N}_k \}$ on system $B$, the largest eigenvalue of the operator
$$
\sum_k p_k {\rm Tr} \left( \rho_k \widetilde{N}_k \right) \rho_k
$$
is bounded by $\frac{1}{2} p_H = \frac{1}{4} \left(1+ \frac{\sqrt{3}}{2} \right)$. The proof that this holds is straight-forward but needs a few steps, and the details are given in appendix \ref{trinedetails}.

The probability of correctly identifying the state using the optimal sequential measurement is given by
\[
{\rm P_{corr}^{seq} } = {\rm Tr}( \Gamma^A ) = p_H = \frac{1}{2} \left( 1+ \frac{\sqrt{3}}{2} \right) \simeq 0.933.
\]

\subsection{Comparison of global and sequential schemes}
For comparison we recall the globally optimal measurement strategy, also discussed in \cite{Chitambar13}. Recall the double trine ensemble satisfies $\ket{\psi_i} \ket{\psi_i} = (U \otimes U)^i \ket{\psi_0} \ket{\psi_0}$, where $U$ is a rotation of $\frac{2 \pi}{3}$ around the $z$-axis in the Bloch sphere. For sets with such symmetry the optimal measurement was shown by Ban \emph{et al} to be given by the so-called square-root measurement \cite{Ban97} (also known as the ``pretty-good measurement'' \cite{Hausladen94}). In this case, the optimal measurement corresponds to a projective measurement, with operators $\{ \Pi_j = \ket{\Phi_j} \bra{\Phi_j} \}$, where
\begin{eqnarray}
\ket{\Phi_j} &=& \frac{1}{\sqrt{3}} \left( \ket{0} \ket{0} +e^{2 \pi j i/3} \frac{1}{\sqrt2}(\ket{01} + \ket{10}) \right. \nonumber \\
&& \left. + e^{4 \pi j i/3} \ket{1} \ket{1} \right). \label{Phi}
\end{eqnarray}
The probability of correctly identifying the state is 
\[
{\rm P_{corr}^{glob}} = \frac{1}{2} + \frac{\sqrt{2}}{3} \simeq 0.971.
\]
Note that the probability of identifying the state correctly achieved by the optimal sequential measurement is greater than $96\%$ of that achieved by the optimal global measurement. In systems where joint measurement is technologically challenging it is thus perhaps difficult to argue that the additional experimental effort is merited by the improvement in performance in this case.

We comment finally on the optimal sequential measurement as an approximation to the optimal global measurement. For the optimal sequential measurement, given above, we obtain
\begin{eqnarray}
\pi_0 &=& \frac{2}{3} \left( \ket{\psi_1^\perp} \bra{\psi_1^\perp} \otimes \ket{\phi_{0|1}} \bra{\phi_{0|1}} \right. \nonumber \\
&&  \left. +\ket{\psi_2^\perp} \bra{\psi_2^\perp} \otimes \ket{\phi_{0|2}} \bra{\phi_{0|2}} \right) \nonumber \\
\pi_1 &=& (U \otimes U) \pi_0 (U \otimes U)^\dagger \label{pi} \\
\pi_2 &=& (U \otimes U)^2 \pi_0 ((U \otimes U)^\dagger)^2 \nonumber
\end{eqnarray}
Considering $\pi_0$, after a little algebra we find
\begin{eqnarray*}
\ket{\psi_1^\perp} \otimes \ket{\phi_{0|1}} &=& \frac{1}{2} e^{- \pi i/12} \left[ \sqrt{1+2 \cos^2 \frac{\pi}{12}} \ket{\alpha_0} \right. \\
&& \left. + i \sqrt{1 + 2\sin^2 \frac{\pi}{12}} \ket{\beta_0} \right] \\
\ket{\psi_2^\perp} \otimes \ket{\phi_{0|2}} &=& \frac{1}{2} e^{\pi i/12} \left[ \sqrt{1+2 \cos^2 \frac{\pi}{12}} \ket{\alpha_0} \right. \\
&& \left. - i \sqrt{1 + 2\sin^2 \frac{\pi}{12}} \ket{\beta_0} \right]
\end{eqnarray*}
where
\begin{eqnarray*}
\ket{\alpha_0} &=& \left(1+2 \cos^2 \frac{\pi}{12} \right)^{-1/2} \left( \cos \frac{\pi}{12} \ket{00} \right. \\
&& \left. + \frac{1}{\sqrt2} \left( \ket{01} + \ket{10} \right) + \cos \frac{\pi}{12} \ket{11} \right) \\
\ket{\beta_0} &=& \left(1 + 2\sin^2 \frac{\pi}{12} \right)^{-1/2} \left( \sin \frac{\pi}{12} \ket{00} \right. \\
&& \left. + \frac{1}{\sqrt2} \left( \ket{01} - \ket{10} \right) - \sin \frac{\pi}{12} \ket{11} \right).
\end{eqnarray*}
Thus we can write
\begin{eqnarray*}
\pi_0 &=& \frac{1}{3} \left(1+2 \cos^2 \frac{\pi}{12} \right) \ket{\alpha_0} \bra{\alpha_0} \\
&& + \frac{1}{3} \left(1+2 \sin^2 \frac{\pi}{12} \right) \ket{\beta_0} \bra{\beta_0} \\
&=& \frac{1}{3} \left( 2+\frac{\sqrt{3}}{2} \right) \ket{\alpha_0} \bra{\alpha_0} \\
&& + \frac{1}{3} \left( 2-\frac{\sqrt{3}}{2} \right) \ket{\beta_0} \bra{\beta_0}.
\end{eqnarray*}
Note that $\braket{\alpha_0}{\beta_0} =0$, and hence this is the eigendecomposition of the operator. We further note that $\ket{\beta_0}$ is orthogonal to the signal state $\ket{\psi_0} \ket{\psi_0}$ and thus does not contribute to the probability of identifying the state. The remaining eigenvector $\ket{\alpha_0}$ is an approximation to $\ket{\Phi_0}$, the state onto which the optimal global measurement projects; an amazingly good one in fact: it turns out $ | \braket{\alpha_0}{\Phi_0} |^2 = 0.9997$. Due to the weighting factor, the overlap between $\ket{\Phi_0}$ and $\pi_0$ is given by $\bra{\Phi_0} \pi_0 \ket{\Phi_0} = \frac{1}{3} \left(2+\frac{\sqrt{3}}{2} \right) | \braket{\alpha_0}{\Phi_0} |^2 = 0.9551$.

The state $\ket{\Phi_0}$ is thus very close to a superposition of $\ket{\psi_1^\perp} \otimes \ket{\phi_{0|1}}$ and $\ket{\psi_2^\perp} \otimes \ket{\phi_{0|2}}$, with appropriate normalisation:
\begin{eqnarray*}
\ket{\Phi_0} \simeq \ket{\alpha_0} &=& \left( 1+2 \cos^2 \frac{\pi}{12} \right)^{-1/2} \left( e^{\pi i/12} \ket{\psi_1^\perp} \otimes \ket{\phi_{0|1}} \right. \\
&& \left. + e^{- \pi i/12} \ket{\psi_2^\perp} \otimes \ket{\phi_{0|2}} \right).
\end{eqnarray*}
The optimal sequential measurement, on the other hand, is formed from a \emph{mixture} of projectors onto these same states. It gives additional information -- one state is ruled out with certainty -- at the expense of a slightly lower probability of success.

\subsection{A non-optimal sequential measurement}
The example of the trine states is further illuminating, as there exists another measurement strategy which satisfies both necessary conditions (\ref{Nopt}) and (\ref{Mopt}), but which is not an optimal strategy, thus demonstrating that these two conditions, when taken together, are not sufficient to define the optimal measurement. This strategy is to perform the optimal minimum error measurement at each step, with Bayesian update of the probabilities in between measurements. Note that such a strategy is known to be optimal (in fact performs as well as the best joint measurement) for a different set of states - the case of just two pure states \cite{Brody96,Acin05}. For the trine states, the measurement is as follows: $\{ M_j \}$ is the optimal one-copy minimum error measurement, which consists of weighted projectors onto the trine states themselves \cite{Helstrom76,Ban97}, $M_j = \frac{2}{3} \ket{\psi_j} \bra{\psi_j}$. Note that for the trine states $|\braket{\psi_i}{\psi_j}|^2 = \frac{1}{4} \left(1 + 3 \delta_{ij} \right)$, and thus the updated priors upon obtaining outcome $j$ are, using equation (\ref{posteriori}):
$$
p_{i|j} = \frac{\frac{2}{3} |\braket{\psi_i}{\psi_j}|^2}{\frac{2}{3} \sum_k |\braket{\psi_k}{\psi_j}|^2} = \frac{1}{6} + \frac{1}{2} \delta_{ij}
$$
For each $j$, the states with these probabilities have so-called mirror symmetry -- the set is invariant under reflection about $\ket{\psi_j}$. For such a set, the minimum error problem was considered by Andersson et al \cite{Andersson02}: using their results we find for $j=0$ the optimal measurement is of the form:
\begin{eqnarray*}
N_{0|0} &=& (1-a^2) \ket{\psi_0} \bra{\psi_0}, \\
N_{1|0} &=& \frac{1}{2} \left( a \ket{\psi_0} - i \ket{\psi_0^\perp} \right) \left( a \bra{\psi_0} + i \bra{\psi_0^\perp} \right), \\
N_{2|0} &=& \frac{1}{2} \left( a \ket{\psi_0} + i \ket{\psi_0^\perp} \right) \left( a \bra{\psi_0} - i \bra{\psi_0^\perp} \right),
\end{eqnarray*}
where $a$ depends on the geometry of the set and the prior probabilities \cite{Andersson02}: for our case we find $a = \frac{1}{5 \sqrt3}$. The optimal measurements for $j=1,2$ are obtained by symmetry $\{ N_{i|j} = U^j N_{i|0} (U^j)^\dagger \}$. Note that condition (\ref{Nopt}) is satisfied by construction. Turning to condition (\ref{Mopt}), we find
\begin{eqnarray*}
{\rm Tr} \left( \rho_0 N_{0|0} \right) &=& \frac{74}{75}, \\
{\rm Tr} \left( \rho_1 N_{1|0} \right) &=& {\rm Tr} \left( \rho_2 N_{2|0} \right) = \frac{32}{75},
\end{eqnarray*}
with analogous results for $j=1,2$. Concisely, ${\rm Tr} \left( \rho_i N_{i|j} \right) = \frac{32}{75} + \frac{42}{75} \delta_{ij}$. Finally, we can calculate $\Gamma^A$:
\begin{eqnarray*}
\Gamma^A &=& \sum_{i,j} p_i {\rm Tr} \left( \rho_i N_{i|j} \right) \rho_i M_j \\
&=& \sum_{i,j} \frac{1}{3} \left( \frac{32}{75} + \frac{42}{75} \delta_{ij} \right) \left( \ket{\psi_i} \bra{\psi_i} \right) \left( \frac{2}{3} \ket{\psi_j} \bra{\psi_j} \right) \\
&=& \frac{30}{75} \mathbbmss{1} = \frac{2}{5} \mathbbmss{1}.
\end{eqnarray*}
For $c_j \sigma_j$ we obtain:
\begin{eqnarray*}
\sum_i p_i {\rm Tr} \left( \rho_i N_{i|j} \right) \rho_i^A &=& \frac{1}{3} \sum_i \left( \frac{32}{75} + \frac{42}{75} \delta_{ij} \right) \rho_i \\
&=& \frac{16}{75} \mathbbmss{1} + \frac{14}{75} \rho_j \\
&=& \frac{2}{5} \ket{\psi_j} \bra{\psi_j} + \frac{16}{75} \ket{\psi_j^\perp} \bra{\psi_j^\perp}
\end{eqnarray*}
from which it is clear that condition (\ref{Mopt}) is satisfied for each $j$.

An analogous situation arises in state discrimination maximising the mutual information between sender and receiver - a necessary but not sufficient condition is known, and for the example of the trine states, is satisfied by both the trine measurement, which is not optimal \cite{Ban97}, and the anti-trine measurement, which is optimal \cite{Sasaki99}. We finally note that the probability of correctly identifying the state using this scheme, ${\rm Tr}(\Gamma^A) = \frac{4}{5}$, is considerably worse than that given by the optimal sequential measurement from above.

\section{General bi-partite case}
\subsection{Necessary and sufficient conditions}
Above, for simplicity, we confined our discussion of optimal sequential measurement strategies to the case of two-copy state discrimination. The conditions obtained however are easily extended to the general bi-partite case. Suppose therefore we are provided with a bi-partite state drawn from a known set $\{ \rho_i^{AB} \}$, with known \emph{a priori} probabilitites $\{ p_i \}$. If our measurement strategy is restricted to sequential measurements on each subsystem, with feed-forward, what is the best measurement to make? The allowed measurements on the joint $AB$ system are again described by POVMs of the form $\{ \pi_i = \sum_j M_j^A \otimes N_{i|j}^B \}$, and the probability of correctly identifying the state is expressed:
$$
{\rm P_{corr}} = \sum_{ij} p_i {\rm Tr}_{AB} \left( \rho_i^{AB} M_j^A \otimes N_{i|j}^B \right).
$$
Following the same reasoning as in Section III, the necessary conditions eqns (\ref{Nopt}), (\ref{Mopt}) become:
\begin{eqnarray*}
\sum_i p_i {\rm Tr}_A \left( \rho_i^{AB} M_j \right) N_{i|j} - p_k {\rm Tr}_A \left( \rho_k^{AB} M_j \right) & \geq & 0, \\
\sum_{i,j} p_i {\rm Tr}_B \left( \rho_i^{AB} N_{i|j} \right) M_j - \sum_i p_i {\rm Tr}_B \left( \rho_i^{AB} N_{i|k} \right) & \geq & 0,
\end{eqnarray*} 
with the following interpretation: given a measurement $M_j^A$ on system $A$, $\{ N_{i|j}^B \}$ must be optimal for discriminating the updated states $\sigma_{i|j}^B$, occuring with probabilities $p_{i|j}$:
\begin{eqnarray*}
\sigma_{i|j}^B &=& \frac{{\rm Tr}_A \left( \rho_i^{AB} M_j^A \right)}{{\rm Tr}_{AB} \left( \rho_i^{AB} M_j^A \right)}, \\
p_{i|j} &=& \frac{p_i {\rm Tr}_{AB} \left( \rho_i^{AB} M_j^A \right)}{\sum_k p_k {\rm Tr}_{AB} \left( \rho_k^{AB} M_j^A \right)}.
\end{eqnarray*}

Similarly, given measurements $\{ \{ N_{i|j}^B \} \}$ on system $B$, $\{ M_j^A \}$ must be optimal for discriminating the states $\sigma_j^A$, occuring with probabilities $q_j$:
\begin{eqnarray*}
\sigma_j^B &=& \frac{\sum_i p_i {\rm Tr}_B \left(\rho_i^{AB} N_{i|j}^B \right)}{\sum_k p_k {\rm Tr}_{AB} \left( \rho_k^{AB} N_{k|j}^B \right)}, \\
q_j &=& \frac{\sum_i p_i {\rm Tr}_{AB} \left( \rho_i^{AB} N_{i|j}^B \right)}{\sum_l \sum_k p_k {\rm Tr}_{AB} \left( \rho_k^{AB} N_{k|l}^B \right)}.
\end{eqnarray*}

Finally, following the same argument as in Section III, the necessary and sufficient condition for optimality of $\{ \pi_j = \sum_j M_j \otimes N_{i|j} \}$ for discriminating the general bipartite states $\{ \rho_i^{AB} \}$ becomes:
\begin{equation}
\sum_{i,j} p_i {\rm Tr}_B \left( \rho_i^{AB} N_{i|j} \right) M_j - \sum_k p_k {\rm Tr}_B \left( \rho_k^{AB} \widetilde{N}_{k} \right) \geq 0, \label{generalcond}
\end{equation}
where $\{ \widetilde{N}_k \}$ is any physically allowed measurement on system B.

\subsection{Example: Three Bell states}
As an example of the general case, we consider the simple case of discriminating between three Bell states $\rho_i^{AB} = \ket{\Psi_i} \bra{\Psi_i}$:
\begin{eqnarray*}
\ket{\Psi_0} &=& \frac{1}{\sqrt2} \left( \ket{0} \ket{0} + \ket{1} \ket{1} \right), \\
\ket{\Psi_1} &=& \frac{1}{\sqrt2} \left( \ket{0} \ket{1} + \ket{1} \ket{0} \right), \\
\ket{\Psi_2} &=& \frac{1}{\sqrt2} \left( \ket{0} \ket{0} - \ket{1} \ket{1} \right),
\end{eqnarray*}
occurring with equal probabilities $p_i = \frac{1}{3}$. Although perfect discrimination between any two Bell states is possible by only local measurements and feed-forward (for example, to distinguish between $\ket{\Psi_0}$ and $\ket{\Psi_1}$ one need only measure both systems in the $\{ \ket{0}, \ket{1} \}$ basis and look at the correlations between outcomes) it is known that for more than two states this is no longer possible \cite{Ghosh01,Walgate02}. To distinguish between all three states, one strategy is to simply perform the measurement that perfectly distinguishes any two states, and never identify the third. We show that this strategy is optimal in terms of minimising the probability of error.

Consider therefore the measurement:
\begin{eqnarray*}
M_0 = \ket{0} \bra{0}, & \; & M_1 = \ket{1} \bra{1}, \\
N_{0|0} = \ket{0} \bra{0}, & \; & N_{1|0} = \ket{1} \bra{1}, \; N_{2|0} = 0, \\
N_{0|1} = \ket{1} \bra{1}, & \; & N_{1|1} = \ket{0} \bra{0}, \; N_{2|1} = 0,
\end{eqnarray*}
that is, both Alice and Bob measure in the $\{ \ket{0}, \ket{1} \}$ basis. Bob takes outcome `$0$' to indicate state $\ket{\Psi_0}$, and outcome `$1$' to indicate state $\ket{\Psi_1}$. State $\ket{\Psi_2}$ is never identified.

It is useful to rewrite equation (\ref{generalcond}) as follows:
\[
\Gamma^A - \widetilde{c} \, \widetilde{\sigma} \geq 0
\]
where
\begin{eqnarray*}
\Gamma^A &=& \sum_{i,j} p_i {\rm Tr}_B \left( \rho_i^{AB} N_{i|j} \right) M_j, \\
\widetilde{c} &=& \sum_k p_k {\rm Tr}_{AB} \left( \rho_k^{AB} \widetilde{N_k} \right), \\
\widetilde{\sigma} &=& \frac{1}{\widetilde{c}}\sum_k p_k {\rm Tr}_B \left( \rho_k^{AB} \widetilde{N}_k \right).
\end{eqnarray*}

Note that $\widetilde{\sigma}$ is a density operator. Further, it is straight-forward to show that
\[
\Gamma^A = \sum_{i,j} p_i {\rm Tr}_B \left( \rho_i^{AB} N_{i|j} \right) M_j = \frac{1}{3} \mathbbmss{1}^A
\]
while
\begin{eqnarray*}
\widetilde{c} &=& \frac{1}{3} \sum_k {\rm Tr}_B \left( {\rm Tr}_A(\rho_k^{AB}) \widetilde{N}_k \right) \\
&=& \frac{1}{3} \sum_k {\rm Tr}_B \left( \left( \frac{1}{2} \mathbbmss{1}^B \right) \widetilde{N}_k \right) \\
&=& \frac{1}{3},
\end{eqnarray*}
where the second line follows as the reduced density operator for system B in all cases is proportional to the identity $\mathbbmss{1}^B$, and the last line follows from the POVM condition $\sum \widetilde{N}_k = \mathbbmss{1}^B$. Thus the condition (\ref{generalcond}) becomes
\[
\mathbbmss{1} - \widetilde{\sigma} \geq 0,
\]
which is true for any arbitrary density operator $\widetilde{\sigma}$. Thus $\{ M_j \otimes N_{i|j} \}$ is an optimal measurement among sequential strategies for discrimination of the three Bell states. A similar approach could be taken to the problem of discriminating all four Bell states, but this complicates the analysis without changing the basic conclusions.

\subsection{Example: Domino states}
\begin{figure}[h!]
\begin{center}
\includegraphics[width=60mm]{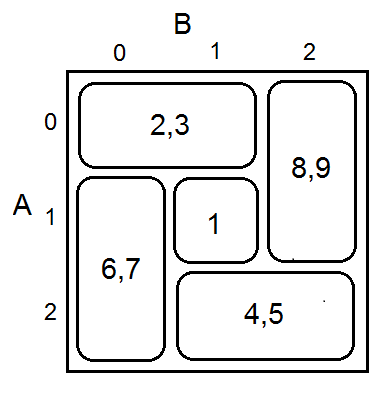}
\end{center}
\caption{Graphical representation of the domino states as domino tiles.}
\label{DominoFig}
\end{figure}
The domino states are an orthonormal basis of a two qutrit system, composed entirely of product states, which nevertheless are not perfectly distinguishable by local measurement and classical communication (LOCC) \cite{Bennett99}. These were the first set of states for which it was shown explicitly that information encoded in product states is not all available through local measurements. The set of states is given by:
\begin{eqnarray*}
\ket{\psi_1} &=& \ket{1} \ket{1}, \\
\ket{\psi_2} &=& \ket{0} \ket{0+1}, \\
\ket{\psi_3} &=& \ket{0} \ket{0-1}, \\
\ket{\psi_4} &=& \ket{2} \ket{1+2}, \\
\ket{\psi_5} &=& \ket{2} \ket{1-2}, \\
\ket{\psi_6} &=& \ket{1+2} \ket{0}, \\
\ket{\psi_7} &=& \ket{1-2} \ket{0}, \\
\ket{\psi_8} &=& \ket{0+1} \ket{2}, \\
\ket{\psi_9} &=& \ket{0-1} \ket{2},
\end{eqnarray*}
where $\ket{i \pm j} = \frac{1}{\sqrt2} \left( \ket{i} \pm \ket{j} \right)$. A useful graphical representation of the states is given in Fig. \ref{DominoFig}. As the states are orthonormal, they are perfectly distinguishable by a joint measurement. While it is known that any local measurement strategy, possibly consisting of many rounds of measurement and classical communication, cannot achieve perfect discrimination, in practice known lower bounds are really very small -- the information deficit of any local protocol is at least $5.31 \times 10^{-6}$ (compared to an achievable value of $\log_2 9 = 3.17$ bits) \cite{Bennett99}, while the probability of error is at least $1.9 \times 10^{-8}$ \cite{Childs13}. Although this is of theoretical interest, if these bounds are achievable it is fair to say that for all practical purposes the states are distinguishable by a local strategy, albeit one with many rounds of communication.

\subsubsection{A candidate sequential measurement}
We recently gave the achievable probability of correct discrimination using sequential measurements for these states \cite{Domino17}. Here we give an alternative proof, which uses the conditions introduced earlier. Motivated by the double-copy trine example, in which the best strategy rules out one state in the first step and discriminates optimally between the remaining two in the second step, we consider strategies that partition the allowed states into subsets in the first step and discriminate between the states within a subset in the second step. Note that amongst the domino states there are a total of 8 subsets that are perfectly distinguishable on the second system alone. These are:
\begin{eqnarray*}
\mathcal{S}_1 &=& \{ \ket{\psi_1}, \ket{\psi_6}, \ket{\psi_8} \} \\
\mathcal{S}_2 &=& \{ \ket{\psi_1}, \ket{\psi_6}, \ket{\psi_9} \} \\
\mathcal{S}_3 &=& \{ \ket{\psi_1}, \ket{\psi_7}, \ket{\psi_8} \} \\
\mathcal{S}_4 &=& \{ \ket{\psi_1}, \ket{\psi_7}, \ket{\psi_9} \} \\
\mathcal{S}_5 &=& \{ \ket{\psi_2}, \ket{\psi_3}, \ket{\psi_8} \} \\
\mathcal{S}_6 &=& \{ \ket{\psi_2}, \ket{\psi_3}, \ket{\psi_9} \} \\
\mathcal{S}_7 &=& \{ \ket{\psi_4}, \ket{\psi_5}, \ket{\psi_6} \} \\
\mathcal{S}_8 &=& \{ \ket{\psi_4}, \ket{\psi_5}, \ket{\psi_7} \}
\end{eqnarray*}
We therefore begin with the conjecture that the best measurement for discriminating the domino states optimally assigns the state to one of these subsets at the first step, and discriminates perfectly between the remaining 3 states in the second step. Thus the only error is introduced in the first step. Denoting now $\ket{\psi_i} = \ket{\phi_i} \ket{\chi_i}$, the probability of identifying the correct state is
\begin{eqnarray*}
{\rm P}_{\rm corr} &=& \sum_{ij} p_i \bra{\phi_i} M_j \ket{\phi_i} \bra{\chi_i} N_{i|j} \ket{\chi_i} \\
&=& \frac{1}{9} \sum_j \sum_{i \in \mathcal{I}_j} \bra{\phi_i} M_j \ket{\phi_i} \\
&=& \frac{8}{3} \sum_j \frac{1}{8} {\rm Tr} \left( \frac{1}{3} \sum_{i \in \mathcal{I}_j} \ket{\phi_i} \bra{\phi_i} M_j \right) 
\end{eqnarray*}
where in the second line we have used the fact that $\bra{\chi_i} N_{i|j} \ket{\chi_i} = 1$ if $\ket{\chi_i}$ is in the subset $\mathcal{S}_j$, and is zero otherwise, and we have defined the index set $\mathcal{I}_j$ such that $i \in \mathcal{I}_j$ if $\ket{\psi_i} \in \mathcal{S}_j$. Thus $\{ M_j \}$ is the optimal measurement for discriminating the states 
\begin{equation}
\rho_j = \frac{1}{3} \sum_{i \in \mathcal{I}_j} \ket{\phi_i} \bra{\phi_i},
\label{rhoj}
\end{equation}
formed by taking an equal mixture of states in the subsets $\{ \mathcal{S}_j \}$, and occuring with equal probabilities $p_j = \frac{1}{8}$.

\subsubsection{Optimal subset discrimination}
The optimal measurement $\{ M_j \}$ and probability of success may be derived analytically, and are given in Appendix \ref{domino}. Here we discuss some of the symmetries of the states which allow us to simplify the problem, before proving that this measurement scheme results in an optimal sequential strategy.

The states $\left\{ \rho_j ={\rm Tr}_B \left( \sum_{i \in \mathcal{I}_j} \ket{\psi_i} \bra{\psi_i} \right) \right\}$  have a lot of symmetry, in particular the set is invariant under the unitaries:
\begin{eqnarray}
U_0 &=& - \ket{0} \bra{0} + \ket{1} \bra{1} + \ket{2} \bra{2}, \nonumber \\
U_1 &=& \ket{0} \bra{2} + \ket{1} \bra{1} + \ket{2} \bra{0}, \label{U01}
\end{eqnarray}
which simply permute the states within the set. To make this point clear, it is useful to list the states explicitly:
\begin{eqnarray}
\rho_1 &=& \frac{1}{3} {\rm Tr}_B \left( \ket{\psi_1} \bra{\psi_1} + \ket{\psi_6} \bra{\psi_6} + \ket{\psi_8} \bra{\psi_8} \right) \nonumber \\
&=& \frac{1}{3} \left( \ket{1} \bra{1} + \ket{1+2} \bra{1+2} + \ket{0+1} \bra{0+1} \right), \nonumber \\
\rho_2 &=& \frac{1}{3} \left( \ket{1} \bra{1} + \ket{1+2} \bra{1+2} + \ket{0-1} \bra{0-1} \right), \nonumber \\
\rho_3 &=& \frac{1}{3} \left( \ket{1} \bra{1} + \ket{1-2} \bra{1-2} + \ket{0+1} \bra{0+1} \right), \nonumber  \\
\rho_4 &=& \frac{1}{3} \left( \ket{1} \bra{1} + \ket{1-2} \bra{1-2} + \ket{0-1} \bra{0-1} \right), \nonumber  \\
\rho_5 &=& \frac{2}{3} \ket{0} \bra{0} + \frac{1}{3} \ket{0+1} \bra{0+1}, \nonumber \\
\rho_6 &=& \frac{2}{3} \ket{0} \bra{0} + \frac{1}{3} \ket{0-1} \bra{0-1}, \nonumber \\
\rho_7 &=& \frac{2}{3} \ket{2} \bra{2} + \frac{1}{3} \ket{1+2} \bra{1+2}, \nonumber \\
\rho_8 &=& \frac{2}{3} \ket{2} \bra{2} + \frac{1}{3} \ket{1-2} \bra{1-2}. \label{rho}
\end{eqnarray}
Clearly, applying either $U_0$ or $U_1$ to any of the states in the set simply results in another state from the same set. $U_k \rho_j U_k^\dagger = \rho_{\sigma_k(j)}$ for some permutation $\sigma_k(j)$. We may therefore expect the optimal measurement $\{ M_j \}$ to have the same symmetries -- indeed for any measurement $\{ M_j \}$, we can construct another measurement $\{ M_{\sigma_k(j)}^\prime = U_k M_j U_k^\dagger \}$ which has the same probability of success. As the figure of merit is linear, a probabilistic mixture of such strategies $\{ M_j^{\prime \prime} = \frac{1}{2} \left(M_j^\prime +M_j \right) \}$ achieves the same probability of success, and further has the symmetry property $ U_k M_j^{\prime \prime} U_k^\dagger = M_{\sigma_k(j)}^{\prime \prime}$. Thus we can restrict attention to measurements which have the same symmetry properties as the states. 

Turning now to the operator
\begin{eqnarray}
\Gamma^A &=& \sum_{ij} p_i \bra{\chi_i} N_{i|j} \ket{\chi_i} \left( \ket{\phi_i} \bra{\phi_i} M_j \right) \nonumber \\
&=& \frac{1}{9} \sum_j \sum_{i \in \mathcal{I}_j} \ket{\phi_i} \bra{\phi_i} M_j \nonumber \\
&=& \frac{8}{3} \sum_j \frac{1}{8} \rho_j M_j,
\label{GammaA}
\end{eqnarray}
we further note that
\begin{eqnarray*}
U_k \Gamma^A U_k^\dagger &=& \frac{8}{3} \sum_j \frac{1}{8} U_k \rho_j M_j U_k^\dagger \\
&=& \frac{8}{3} \sum_j \frac{1}{8} \left( U_k \rho_j U_k^\dagger \right) \left( U_k M_j U_k^\dagger \right) \\
&=& \frac{8}{3} \sum_j \frac{1}{8} \rho_{\sigma_k(j)} M_{\sigma_k(j)} = \Gamma^A.
\end{eqnarray*}
Thus $\Gamma^A$ is invariant under both $U_0$, and $U_1$, and it follows that it must have the form:
\begin{eqnarray}
\Gamma^A &=& \frac{1}{3} \sum_j \rho_j M_j, \nonumber \\
&=& p \left( \ket{0} \bra{0} + \ket{2} \bra{2} \right) + q \ket{1} \bra{1}.
\label{GammaA2}
\end{eqnarray}
In Appendix \ref{domino} we show that $p=\frac{1}{9} \frac{1}{12} (17 + \sqrt{7} \sqrt{31}) \simeq 0.294 $, $q=\frac{1}{9} \frac{1}{16}(21 + \sqrt{7} \sqrt{31}) \simeq 0.248$, giving a probability of correctly identifying the state of
\[
{\rm P}_{\rm corr} = 2 p + q \simeq 0.836.
\]

\subsubsection{An optimal sequential strategy}
We can now use condition (\ref{generalcond}) to prove that this is the best possible sequential measurement. Again, we outline here some of the symmetry arguments that allow us to simplify the problem, and give the remaining details in Appendix \ref{domino}. Eqn. \ref{generalcond} becomes
\begin{eqnarray*}
\Gamma^A - \frac{1}{9} \sum_k \bra{\chi_k} \widetilde{N}_k \ket{\chi_k} \ket{\phi_k} \bra{\phi_k} & \geq & 0.
\end{eqnarray*}
It is useful to define the non-normalized state
\begin{eqnarray}
\widetilde{\sigma} &=& \sum_k \bra{\chi_k} \widetilde{N}_k \ket{\chi_k} \ket{\phi_k} \bra{\phi_k} \nonumber \\
&=& \bra{1} \widetilde{N}_1 \ket{1} \ket{1} \bra{1} \nonumber \\
&& + \left( \bra{0+1} \widetilde{N}_2 \ket{0+1} + \bra{0-1} \widetilde{N}_3 \ket{0-1} \right) \ket{0} \bra{0} \nonumber \\
&& + \left( \bra{1+2} \widetilde{N}_4 \ket{1+2} + \bra{1-2} \widetilde{N}_5 \ket{1-2} \right) \ket{2} \bra{2} \nonumber \\
&& + \bra{0} \widetilde{N}_6 \ket{0} \ket{1+2} \bra{1+2} + \bra{0} \widetilde{N}_7 \ket{0} \ket{1-2} \bra{1-2} \nonumber \\
&& + \bra{2} \widetilde{N}_8 \ket{2} \ket{0+1} \bra{0+1} + \bra{2} \widetilde{N}_9 \ket{2} \ket{0-1} \bra{0-1}, \nonumber \\ \label{sigmatilde}
\end{eqnarray}
and we wish to show that $\Gamma^A - \frac{1}{9} \widetilde{\sigma} \geq 0$, for all $\widetilde{\sigma}$ resulting from an allowed measurement $\{ \widetilde{N}_j \}$.

We can again use symmetry properties of the states to considerably restrict the set of measurements we need to consider. Consider this time the unitaries
\begin{eqnarray*}
V_0 &=& - \ket{0} \bra{0} + \ket{1} \bra{1} + \ket{2} \bra{2} \\
V_1 &=& \ket{0} \bra{0} + \ket{1} \bra{1} - \ket{2} \bra{2}
\end{eqnarray*}
Note that the operator $\widetilde{\sigma}$ is unchanged if we make the substitution
\begin{eqnarray*}
\widetilde{N}_j^\prime & = & V_0 \widetilde{N}_j V_0^\dagger, \quad j \neq 2,3 \\
\widetilde{N}_2^\prime & = & V_0 \widetilde{N}_3 V_0^\dagger, \\
\widetilde{N}_3^\prime & = & V_0 \widetilde{N}_2 V_0^\dagger,
\end{eqnarray*}
or indeed under a strategy which is formed from a mixture of such strategies. We can make similar arguments for $V_1$, with the result that we only need consider strategies such that the following combinations of operators are invariant under both $V_0$ and $V_1$, and therefore must be diagonal in the $\{ \ket{0}, \ket{1}, \ket{2} \}$ basis: $\{ \widetilde{N}_1, \widetilde{N}_2 + \widetilde{N}_3, \widetilde{N}_4 + \widetilde{N}_5, \widetilde{N}_6, \widetilde{N}_7, \widetilde{N}_8, \widetilde{N}_9 \}$. Further, amongst such strategies it is clear that in order to maximise the coefficients $\bra{\chi_k} \widetilde{N}_k \ket{\chi_k}$ we only need consider those such that
\begin{eqnarray}
\widetilde{N}_1 &=& c_1 \ket{1} \bra{1}, \nonumber \\
\widetilde{N}_2 &=& c_{2} \left( \cos \alpha \ket{0} + \sin \alpha \ket{1} \right) \left( \cos \alpha \bra{0} + \sin \alpha \ket{1} \right), \nonumber \\
\widetilde{N}_3 &=& c_{2} \left( \cos \alpha \ket{0} - \sin \alpha \ket{1} \right) \left( \cos \alpha \bra{0} - \sin \alpha \ket{1} \right), \nonumber \\
\widetilde{N}_4 &=& c_{4} \left( \cos \beta \ket{2} + \sin \beta \ket{1} \right) \left( \cos \beta \bra{2} + \sin \beta \ket{1} \right), \nonumber \\
\widetilde{N}_5 &=& c_{4} \left( \cos \beta \ket{2} - \sin \beta \ket{1} \right) \left( \cos \beta \bra{2} - \sin \beta \ket{1} \right), \nonumber \\
\widetilde{N}_6 &=& c_6 \ket{0} \bra{0}, \nonumber \\
\widetilde{N}_7 &=& c_7 \ket{0} \bra{0}, \nonumber \\
\widetilde{N}_8 &=& c_8 \ket{2} \bra{2}, \nonumber \\
\widetilde{N}_9 &=& c_9 \ket{2} \bra{2}, \label{N}
\end{eqnarray}
where $ 0 \leq \alpha, \beta \leq \pi/2$ and for constants $c_1, c_2, c_4, c_6, c_7, c_8, c_9 \geq 0$ satisfying
\begin{eqnarray*}
c_1 + 2 c_2 \sin^2 \alpha + 2 c_4 \sin^2 \beta &=& 1, \\
2 c_2 \cos^2 \alpha + c_6 + c_7 &=& 1, \\
2 c_4 \cos^2 \beta + c_8 + c_9 &=& 1.
\end{eqnarray*}
First note that for $\alpha = \beta = \frac{\pi}{4}$ all such measurements are convex combinations of the eight fiducial measurements
\begin{eqnarray*}
&& \{ \widetilde{N}_1 = \ket{1} \bra{1}, \widetilde{N}_i = \ket{0} \bra{0}, \widetilde{N}_j = \ket{2} \bra{2}\}, \\
&& \{ \widetilde{N}_2 = \ket{0+1} \bra{0+1}, \widetilde{N}_3 = \ket{0-1} \bra{0-1}, \widetilde{N}_{j} = \ket{2} \bra{2} \}, \\
&& \{ \widetilde{N}_4 = \ket{1+2} \bra{1+2}, \widetilde{N}_5 = \ket{1-2} \bra{1-2}, \widetilde{N}_{i} = \ket{0} \bra{0} \},
\end{eqnarray*}
where $i=6,7$, $j=8,9$. Each such measurement leads to $\widetilde{\sigma} = 3 \rho_j$ for some $j$. Since for $\rho_j$ defined in eqn \ref{rhoj}, $\Gamma^A - \frac{1}{3} \rho_j \geq 0$ by construction, any convex combination of such measurements satisfies $\Gamma^A - \frac{1}{9} \widetilde{\sigma} \geq 0$, as required.

The remaining cases, in which $\alpha, \beta \neq \frac{\pi}{4}$ must be considered separately. It is straight-forward to show that $\Gamma^A - \frac{1}{9} \widetilde{\sigma} \geq 0$ for these cases also -- details of all possible measurements and the corresponding calculations are given in Appendix \ref{domino}.

Thus we find that the optimal sequential strategy for discriminating between the domino states with equal priors identifies the state correctly with probability $83.6 \%$. As the optimal joint measurement discriminates the states perfectly, there is a significant gap in performance, which could in principle be demonstrated experimentally.

\section{A related discrimination problem}
We conclude by discussing an interpretation of the states $\sigma_j$, defined in equation (\ref{sigma}), and thus a second discrimination problem for which $\{ M_j \otimes N_{i|j} \}$ provides an optimal measurement. For simplicity we refer to the two copy case throughout, but the discussion applies equally to the general bi-partite case. Recall that, in a measurement on system B described by POVM $\{ N_{i|j} \}$, the probability of identifying a given state $\rho_i$ correctly from the set $\{ \rho_k \}$ with prior probabilities $\{ p_k \}$ is given by the joint probability:
$$
{\rm P}(\rho_i, i) = {\rm P}(\rho_i) {\rm P}(i | \rho_i) = p_i {\rm Tr}( \rho_i N_{i|j} ),
$$
while the overall probability of correctly identifying the prepared state is obtained by summing over each possible $i$: ${\rm P} ({\rm corr} | \{N_{i|j} \}) = \sum_i p_i {\rm Tr}(\rho_i N_{i|j})$, where we have chosen the notation to make explicit the dependence on the choice of measurement $\{ N_{i|j} \}$. We can thus interpret the ratio of these as the conditional probability that state $\rho_i$ was prepared given that measurement $\{ N_{i|j} \}$ was performed and the state was identified correctly as a result. Explicitly:
$$
{\rm P}(\rho_i |{\rm corr}, \{N_{i|j} \}) = \frac{p_i {\rm Tr} \left( \rho_i N_{i|j} \right)}{\sum_k p_k {\rm Tr} \left( \rho_k N_{k|j} \right)}.
$$
We thus arrive at an interpretation for the state $\sigma_j^A$ -- it is the state we should assign to system $A$, given knowledge that a measurement $\{ N_{i|j} \}$ performed on system $B$ identified the state correctly (but importantly, \emph{without} knowledge of the particular outcome of measurement on $B$):
$$
\sigma_j^A = \sum_i {\rm P}(\rho_i |{\rm corr}, \{N_{i|j} \}) \rho_i^A.
$$
Similarly, consider the probabilities $q_j$: if a measurement is chosen from the set $\{ \{ N_{i|j} \} \}$ with equal probabilities $1/n$, then $q_j$ represents the conditional probability that the measurement used was $\{ N_{i|j} \}$, given the knowledge that the state was correctly identified:
\begin{eqnarray*}
q_j = \frac{c_j}{\sum_k c_k} &=& \frac{\frac{1}{n} {\rm P}({\rm corr}| \{ N_{i|j} \})}{\frac{1}{n} \sum_k {\rm P}({\rm corr}| \{ N_{i|k} \})} \\
&=& {\rm P}(\{ N_{i|j} \} | {\rm corr}).
\end{eqnarray*}
$\{ M_j \}$ is thus the optimal measurement for discriminating the residual states on system $A$, given that a measurement chosen from the set $\{ \{ N_{i|j} \} \}$ was performed on system $B$, and the state of system $B$ was correctly identified as a result. $\{ M_j \}$ is the measurement that allows us to optimally guess which measurement was performed on $B$.

Suppose further that another measurement $\{  \widetilde{N}_i \}$ is added to the set of possible measurements $\{ N_{i|j} \}$, on $B$. If we now define
\begin{eqnarray*}
\widetilde{\sigma}^A &=& \sum_i {\rm P}(\rho_i |{\rm corr}, \{\widetilde{N}_i \}) \rho_i^A, \\
q_j^\prime &=& \frac{\frac{1}{n+1} {\rm P}({\rm corr}| \{ N_{i|j} \})}{\frac{1}{n+1} \left( \sum_k {\rm P}({\rm corr}| \{ N_{i|k} \}) +  {\rm P}({\rm corr}| \{  \widetilde{N}_{i} \}) \right)} \\
&=& {\rm P}(\{ N_{i|j} \} | {\rm corr}), \\
\widetilde{q} &=& \frac{\frac{1}{n+1} {\rm P}({\rm corr}| \{ \widetilde{N}_i \})}{\frac{1}{n+1} \left( \sum_k {\rm P}({\rm corr}| \{ N_{i|k} \}) +  {\rm P}({\rm corr}| \{  \widetilde{N}_{i} \}) \right)} \\
&=& {\rm P}(\{ \widetilde{N}_i \} | {\rm corr}),
\end{eqnarray*}
we can rewrite condition (\ref{seqcond}) as follows:
\[
\sum_{i,j} q_j^\prime \sigma_j^A M_j^A - \widetilde{q} \, \widetilde{\sigma}^A \geq 0,
\]
where $\sigma_j^A$ are defined as before. This tells us that the measurement which discriminates optimally between the states $\{ \sigma_j^A \}$, but which never identifies $\{ \widetilde{\sigma} \}$ remains optimal for the new set. That is, the optimal measurement $\{ M_j \}$ is unchanged by the addition of a new measurement to the set used to measure system $B$.

We thus arrive at a new discrimination problem, seemingly unrelated to our original problem of interest, but for which the measurement $\{ M_j^A \otimes N_{i|j} \}$ also provides an optimal strategy, and for which the conditions (\ref{Mopt}) and (\ref{seqcond}) have a natural interpretation. Consider the following game involving three parties, Alice, Bob, and Claire. Claire prepares two copies of a state $\rho_i$, drawn from a given set $\{ \rho_j \}$ with priors $\{ p_j \}$, and sends one copy to Alice and one to Bob. Claire also sends an index $j$ to Bob (this is classical information and may be sent over a classical channel), chosen with equal probabilities from the set $0 \leq j \leq n-1$, where $n$ is an integer chosen by Alice and Bob, and is the payout they will receive if they win the game. Alice and Bob each perform a measurement on their system, following which Bob must make a guess as to the state $\rho_i$, while Alice must guess the index $j$. Alice and Bob can pre-agree on a strategy, but cannot communicate with one another once the game begins. They win only if both guesses are correct. Their strategy is as follows: Alice always makes the measurement $\{ M_j \}$, while Bob's chosen measurement depends on the index $j$ received from Claire -- given $j$, Bob makes measurement $\{ N_{i|j} \}$. The probability that Alice and Bob win the game is simply
\begin{eqnarray*}
{\rm P(win)} &=& \sum_{i,j} {\rm P}(i) {\rm P}(j) {\rm Tr}(\rho_i N_{i|j}) {\rm Tr} (\rho_i M_j) \\
 &=& \frac{1}{n} \sum_{i,j} p_i {\rm Tr}(\rho_i N_{i|j}) {\rm Tr} (\rho_i M_j),
\end{eqnarray*}
and the expected payout is thus
\[
\langle {\rm Payout} \rangle = n {\rm P(win)} = \sum_{i,j} p_i {\rm Tr}(\rho_i N_{i|j}) {\rm Tr} (\rho_i M_j).
\]
This is the same figure of merit as that given in eqn (\ref{Pcorrseq}). Thus the optimal measurement strategy maximising the expected payout for Alice and Bob is the same as the optimal sequential measurement discriminating the bi-partite states $\{ \rho_i \otimes \rho_i \}$.

\section{Discussion}
We have discussed the problem of extracting classical information from a set of bi-partite states, when the measurement strategy is restricted to sequential measurements of each subsystem, with feed-forward of classical information in between measurements. As this is a physically well-motivated class, it is useful to understand how well it performs compared to the ability to perform arbitrary joint measurements, which in many physical systems is still technologically challenging. We have constructed an analogue of the Helstrom conditions for sequential measurement strategies. Like the Helstrom conditions, it is not obvious how to use this condition to construct an optimal measurement, but we show how for certain examples it is possible to use the condition to prove optimality of a candidate measurement procedure.

Our necessary and sufficient condition for optimality of a given sequential measurement still contains an arbitrary measurement on one subsystem. We have been unable to find a condition which is both necessary and sufficient and requires only the set of states and a candidate measurement. It would certainly be useful to find one, but in the absence of such, given a candidate optimal measurement our condition reduces the complexity of checking optimality from optimising over both systems to just optimising over one. It would also be interesting in the future to extend this analysis to other figures of merit such as those which interpolate between minimum-error and unambiguous discrimination \cite{Chefles98}, or which maximise the success rate of discrimination while allowing for inconclusive results \cite{Herzog2015,Fiurasek2003}.

For the two-copy trine case, the probability of success of the optimal sequential measurement is 96 \% of the value achieved by the optimal global measurement \cite{Chitambar13}. The optimal sequential measurement sometimes rules out one of the states with certainty, thus providing information not given by the optimal global strategy, at the expense of a slightly higher probability of failure. Nonetheless, the difference in performance is arguably too small to motivate experimental implementation of the joint measurement. In fact, although we do not give the details here, we have found that for two copies of any set of symmetric qubit states, the optimal sequential measurement performs almost as well as the optimal joint measurement.

We have given the optimal sequential measurement for discriminating the domino states. This set of orthogonal product states has played an important role in quantifying the difference between local measurements and separable measurements \cite{Bennett99}, but the more practical question of how well a sequential measurement performs seems not to have been considered in the literature until recently \cite{Domino17}. We gave here an alternative proof for the optimal sequential measurement, which reveals a significant gap between the probability of success of discrimination of the optimal sequential measurement (83.6\%) and the optimal global measurement, which discriminates the states perfectly.

We further introduced a complementary discrimination problem, in the form of a three-party game, which requires optimisation of the same figure of merit as our original problem of minimising the error over sequential measurement strategies. This game arises naturally in providing an interpretation for the conditions an optimal measurement must satisfy, and provides a new perspective on the sequential measurement problem.

\acknowledgments{This work was supported by the University of Glasgow College of Science and Engineering (S.C. \& G.W.) and by the Royal Society Research Professorships (S.M.B., Grant No RP150122).}

\appendix
\section{Proof of optimality for the double trine ensemble \label{trinedetails}}
To prove optimality of the sequential measurement scheme given in the text, we wish to show that the largest eigenvalue of
\[
\widetilde{\sigma} = \frac{1}{3} \sum_k {\rm Tr}(\rho_k \widetilde{N}_k) \rho_k
\]
is less than or equal to $\frac{1}{2} p_H = \frac{1}{4} \left( 1+ \frac{\sqrt{3}}{2} \right)$ for any physically allowed measurement $\{ \widetilde{N}_k \}$. We begin by writing the trine states $\rho_j$ in the Bloch sphere representation:
\[
\rho_j = \frac{1}{2} \left( I + \cos \left(\frac{2 \pi j}{3} \right) \sigma_x + \sin \left(\frac{2 \pi j}{3} \right) \sigma_y \right).
\]
Writing $s_k = \frac{1}{3} {\rm Tr} \left( \rho_k \widetilde{N}_k \right)$ we thus obtain
\begin{eqnarray*}
\widetilde{\sigma} &=& \frac{1}{2} ( (s_0+s_1+s_2) I \\
&& + (s_0 -\frac{1}{2} (s_1+s_2)) \sigma_x + \frac{\sqrt{3}}{2} (s_1-s_2) \sigma_y )
\end{eqnarray*}
with eigenvalues
\begin{eqnarray*}
\lambda_\pm &=& \frac{1}{2} ( s_0 + s_1 + s_2) \\
&& \pm \frac{1}{2} \left( \sqrt{\left(s_0 -\frac{1}{2} (s_1+s_2)\right)^2 + \left(\frac{\sqrt{3}}{2} (s_1-s_2)\right)^2} \right) \\
&=& \frac{1}{2} ( s_0 + s_1 + s_2) \\
&& \pm \frac{1}{2} |s_0 +e^{\frac{2 \pi i}{3}} s_1+ e^{-\frac{2 \pi i}{3}} s_2|.
\end{eqnarray*}
Thus it follows that there exists some $\theta$ such that the largest eigenvalue, $\lambda_+$ may be written:
\begin{eqnarray*}
\lambda_+ &=& \frac{1}{2} ( s_0 + s_1 + s_2) \\
&& + \frac{1}{2} e^{i \theta} \left( s_0 +e^{\frac{2 \pi i}{3}} s_1+ e^{-\frac{2 \pi i}{3}} s_2 \right), \\
&=& \frac{1}{2} (1+\cos \theta)  s_0 + \frac{1}{2} \left[1 + \cos \left(\theta+ \frac{2 \pi}{3} \right) \right] s_1 \\
&& + \frac{1}{2} \left[1+\cos \left(\theta - \frac{2 \pi}{3} \right) \right] s_2 \\
&=& \frac{1}{2} \left[ \sum_k q_k {\rm Tr} \left( \rho_k \widetilde{N}_k \right) \right],
\end{eqnarray*}
where in the second equality we use the fact that $\lambda_+$ is real, and in the last line we have substituted for $s_k$, and defined $q_k = \frac{1}{3} \left(1+ \cos \left( \theta + \frac{2 \pi k}{3} \right) \right)$. Each strategy $\{ \widetilde{N}_k \}$ thus defines a $\theta$ such that the above equalities hold. For each such $\theta$, we can find an upper bound for $\lambda_+$ by considering the optimisation problem of discriminating the states $\{ \rho_k \}$ occuring with priors $q_k$:
\[
\lambda_+ \leq \frac{1}{2} {\rm P_{corr}} \left( \{ q_k \rho_k \} \right).
\]
We thus wish to find the optimal strategy $\{ \pi_k \}$ for discriminating the trine states with \emph{a priori} probabilities $\frac{1}{3} \left( 1+ \cos \left( \theta + \frac{2 \pi k}{3} \right) \right)$, ultimately maximising the probability of correctness also over $\theta$. Finally, if this maximum is achievable then we have succeeded in finding the optimal $\lambda_+$.

We first choose, without loss of generality, to consider $q_0 \geq q_1 \geq q_2$. This corresponds to $-\frac{\pi}{3} \leq \theta \leq 0$. We use the strategy in \cite{Chitambar13} of bounding the probability of correctly identifying the state by considering the problem of discriminating $q_0 \rho_0$ from $q_1 \rho_1 + q_2 \rho_2$. As any strategy which discriminates all three states also discriminates these two states, the optimal probability of correctly identifying the state for this problem is greater than or equal to that for discriminating all three states 
\begin{eqnarray*}
{\rm P_{corr}} (\{ q_k \rho_k \}) & \leq & {\rm P_{corr}} (q_0 \rho_0, q_1 \rho_1 + q_2 \rho_2) \\
&=& \frac{1}{2} (1+{\rm Tr}|q_0 \rho_0 -(q_1 \rho_1 + q_2 \rho_2)|)
\end{eqnarray*}
Substituting for $q_k$, $\rho_k$, after a little algebra we obtain:
\begin{eqnarray*}
{\rm P_{corr}} (\{ q_k \rho_k \}) & \leq & \frac{1}{2} \left(1+\frac{2}{3} \sqrt{\left(1+\frac{1}{4} \cos \theta \right)^2 + \left(\frac{3}{4} \sin \theta \right)^2} \right).
\end{eqnarray*}
For $\theta$ in the range $-\frac{\pi}{3} \leq \theta \leq 0$, this is a monotonically decreasing function of $\theta$, thus the maxmimum occurs at the boundary of the allowed domain, at $\theta = -\frac{\pi}{3}$, corresponding to
\begin{eqnarray*}
{\rm P_{corr}} (\{ q_k \rho_k \}) & \leq & \frac{1}{2} \left(1+\frac{\sqrt{3}}{2} \right).
\end{eqnarray*}
Thus we obtain $\lambda_+ \leq \frac{1}{4} \left( 1 + \frac{\sqrt{3}}{2} \right)$, as desired. Further this bound is achievable, by the strategy given by \cite{Chitambar13} and outlined in the main text.

\section{Domino states \label{domino}}
We begin with the optimal measurement $\{ M_j \}$ to discriminate the states $\{ \rho_j \}$ given by eqn (\ref{rho}), occuring with equal probabilities $p_j = p = \frac{1}{8}$. Our strategy is to search for $\Gamma = \frac{1}{8} \sum_j \rho_i M_i$ satisfying $\Gamma - \frac{1}{8} \rho_j \geq 0$, and then use eqn (\ref{altcond}) to find $\{ M_j \}$. In fact, as $\Gamma^A$ defined in eqn (\ref{GammaA}) is simply proportional to $\Gamma$, $\Gamma^A = \frac{8}{3} \Gamma$, and eqn (\ref{altcond}) still holds when multiplied by a constant, we work directly with $\Gamma^A$ and search for $\Gamma^A$, $\{ M_j \}$ satisfying
\begin{eqnarray*}
\Gamma^A - \frac{1}{3} \rho_j & \geq & 0, \\
\left( \Gamma^A - \frac{1}{3} \rho_j \right) M_j &=& 0
\end{eqnarray*}
for all $j$.

Recall, as argued in the main text, that $\Gamma^A$ is invariant under the unitaries $U_0$, $U_1$ from eqn (\ref{U01}), and thus has the form $\Gamma^A = p \left( \ket{0} \bra{0} + \ket{2} \bra{2} \right) + q \ket{1} \bra{1}$. We further note that it suffices to check the inequality for $\rho_1$ and $\rho_5$, as all other states can be obtained from these by application of some combination of $U_0$ and $U_1$, by symmetry it then holds for all $j$. For $\rho_1$, this condition -- expressed in matrix form in the $\{ \ket{0+2}, \ket{1}, \ket{0-2} \}$ basis -- becomes:
\[
\Gamma^A - \frac{1}{3} \rho_1 = \left( \begin{array}{ccc} p - \frac{1}{18} & - \frac{1}{9 \sqrt2} & 0 \\ - \frac{1}{9 \sqrt2} & q - \frac{2}{9} & 0 \\ 0 & 0 & p - \frac{1}{18} \end{array} \right) \geq 0.
\]
Thus we require $p \geq \frac{1}{18}$, $q \geq \frac{2}{9}$ and
\begin{equation}
\left(p-\frac{1}{18} \right) \left( q - \frac{2}{9} \right) - \left(\frac{1}{9 \sqrt2}\right)^2 \geq 0.
\label{ineq1}
\end{equation}
For $\rho_5$, the condition is more conveniently expressed in the $\{ \ket{0}, \ket{1}, \ket{2} \}$ basis, in which it has matrix form:
\[
\Gamma^A - \frac{1}{3} \rho_5 = \left( \begin{array}{ccc} p - \frac{5}{18} & - \frac{1}{18} & 0 \\ - \frac{1}{18} & q - \frac{1}{18} & 0 \\ 0 & 0 & p \end{array} \right) \geq 0,
\]
leading to the additional conditions $p \geq \frac{5}{18}$, and
\begin{equation}
\left(p-\frac{5}{18} \right) \left( q - \frac{1}{18} \right) - \left(\frac{1}{18} \right)^2 \geq 0.
\label{ineq2}
\end{equation}
Equality in (\ref{ineq1}) and (\ref{ineq2}) means that each of the operators $\Gamma^A-\frac{1}{3} \rho_j$ has a zero eigenvalue: $M_j$ is then a weighted projector onto the corresponding eigenvector in each case. Setting each equal to zero therefore and solving for $p$ and $q$ gives
\begin{eqnarray}
p &=& \frac{1}{9} \frac{1}{12} (17 + \sqrt{7} \sqrt{31}) \simeq 0.294 , \nonumber \\
q &=& \frac{1}{9} \frac{1}{16}(21 + \sqrt{7} \sqrt{31}) \simeq 0.248. \label{pq}
\end{eqnarray}
$\{ M_j \}$ are then fixed up to multiplying factors, which may be chosen such that $\sum_j M_j = I$.

We now turn to the proof that the sequential measurement strategy given in the main text is indeed the optimal strategy. We wish to show that $\Gamma^A - \frac{1}{9} \widetilde{\sigma} \geq 0$ for all $\widetilde{\sigma}$, of the form given in eqn (\ref{sigmatilde}), where according to symmetry arguments we only need consider measurements $\{ \widetilde{N}_j \}$ of the form (\ref{N}). Note that \emph{all} measurements of this form may be considered to be probabilistic mixtures of a smaller set of fiducial measurements. An exhaustive list of those that cannot be decomposed as a mixture of other measurements are those in which the following elements are the only non-zero ones:
\begin{eqnarray*}
\{ \widetilde{N}_1, \widetilde{N}_i, \widetilde{N}_j \}, & \; & i = 6,7, j = 8,9, \alpha = \beta = \pi/4, \\
\{ \widetilde{N}_2, \widetilde{N}_3, \widetilde{N}_j \}, && j = 8,9, \alpha = \beta = \pi/4, \\
\{ \widetilde{N}_4, \widetilde{N}_5, \widetilde{N}_i \}, && i = 6,7, \alpha = \beta = \pi/4, \\
\{ \widetilde{N}_1, \widetilde{N}_2, \widetilde{N}_3, \widetilde{N}_j \}, && j = 8,9, \alpha < \pi/4, \\
\{ \widetilde{N}_2, \widetilde{N}_3, \widetilde{N}_i, \widetilde{N}_j \}, && i = 6,7, j = 8,9, \alpha > \pi/4, \\
\{ \widetilde{N}_1, \widetilde{N}_2, \widetilde{N}_3, \widetilde{N}_4, \widetilde{N}_5 \}, && \tan^2 \alpha + \tan^2 \beta < 1, \\
\{ \widetilde{N}_2, \widetilde{N}_3, \widetilde{N}_4, \widetilde{N}_5, \widetilde{N}_i \}, && i = 6,7, \tan^2 \alpha + \tan^2 \beta > 1, \\
\{ \widetilde{N}_2, \widetilde{N}_3, \widetilde{N}_4, \widetilde{N}_5, \widetilde{N}_j \}, && j = 8,9, \tan^2 \alpha + \tan^2 \beta > 1, \\
\{ \widetilde{N}_1, \widetilde{N}_4, \widetilde{N}_5, \widetilde{N}_i \}, && j = 6,7, \beta < \pi/4, \\
\{ \widetilde{N}_4, \widetilde{N}_5, \widetilde{N}_i, \widetilde{N}_j \}, && i = 6,7, j = 8,9, \beta > \pi/4,
\end{eqnarray*}
The first three correspond to the case $\widetilde{\sigma} = 3 \rho_j$ for some $j$, and the operator inequality holds by construction. The last three are obtained from others by applying $U_1$, and so by symmetry we only need check the inequality for the remaining four. We consider the remaining possibilities in turn.

{\bf Case 1}: $\alpha < \frac{\pi}{4}$:

All measurements in which $\{ \widetilde{N}_1, \widetilde{N}_2, \widetilde{N}_3, \widetilde{N}_j \}$, $j=8,9$ are the only non-zero elements may be parametrised in the following way, where $r = \tan^2 \alpha$, $0 \leq r < 1$:
\begin{eqnarray*}
\widetilde{N}_1 &=& (1-r) \ket{1} \bra{1}, \\
\widetilde{N}_2 &=& \frac{1}{2} \left( \ket{0} + \sqrt{r} \ket{1} \right) \left( \bra{0} + \sqrt{r} \bra{1} \right), \\
\widetilde{N}_3 &=& \frac{1}{2} \left( \ket{0} - \sqrt{r} \ket{1} \right) \left( \bra{0} - \sqrt{r} \bra{1} \right), \\
 \widetilde{N}_j &=& \ket{2} \bra{2}.
\end{eqnarray*}
For $j=8$, this leads to 
\begin{eqnarray*}
\widetilde{\sigma} &=& (1-r) \ket{1} \bra{1} + \frac{1}{2} (1+ \sqrt{r} )^2 \ket{0} \bra{0} + \ket{0+1} \bra{0+1} \\
& \leq & (1-r) \ket{1} \bra{1} + (1+r) \ket{0} \bra{0} + \ket{0+1} \bra{0+1},
\end{eqnarray*}
where we have used the inequality $(a+b)^2 \leq 2 (a^2 + b^2)$. Note that the $j=9$ case is obtained by applying $U_0$ to $\widetilde{\sigma}$ and thus by symmetry we only need check the $j=8$ case. For this case, in matrix notation, in the $\{ \ket{0}, \ket{1}, \ket{2} \}$ basis, we obtain
\[
\Gamma^A - \frac{1}{9} \widetilde{\sigma} \geq \left( \begin{array}{ccc} p - \frac{1}{9}(1+r) - \frac{1}{18} & -\frac{1}{18} & 0 \\ -\frac{1}{18} & q - \frac{1}{9}(1-r) - \frac{1}{18} & 0 \\ 0 & 0 & p \end{array} \right).
\]
Note that with the values of $p$ and $q$ calculated in eqn (\ref{pq}), the diagonal elements are all strictly positive, and we further require $\left( p - \frac{1}{9}(1+r) - \frac{1}{18})(q-\frac{1}{9}(1-r)-\frac{1}{18} \right) - \left(\frac{1}{18}\right)^2 \geq 0$. Re-writing the first term in brackets as $p - \frac{5}{18} + \frac{1}{9}(1-r)$ we obtain
\begin{eqnarray*}
&&\left(p-\frac{5}{18}\right)\left(q-\frac{1}{18}\right)-\left(\frac{1}{18}\right)^2 \\
&&+\frac{1}{9}(1-r)(q-\frac{1}{18} -(p-\frac{5}{18}) - \frac{1}{9}(1-r)) \geq  0.
\end{eqnarray*}
The combination of the first two terms is positive according to eqn (\ref{ineq2}), and the last term is also positive, as is readily seen by noting $0< r < 1$ and by explicit substitution for $p$ and $q$ (or as a quick check, note that $p < \frac{6}{18}$, while $q > \frac{2}{9}$). Thus the inequality holds.

{\bf Case 2}: $\alpha > \frac{\pi}{4}$:

All measurements in which $\{ \widetilde{N}_2, \widetilde{N}_3, \widetilde{N}_i, \widetilde{N}_j \}$, $i=6,7$, $j=8,9$ are the only non-zero elements may be parametrised in the following way, where $r = \frac{1}{\tan^2 \alpha}$, $0 \leq r < 1$:
\begin{eqnarray*}
\widetilde{N}_2 &=& \frac{1}{2} \left(\sqrt{r} \ket{0} + \ket{1} \right) \left(\sqrt{r} \bra{0} + \bra{1} \right), \\
\widetilde{N}_3 &=& \frac{1}{2} \left(\sqrt{r} \ket{0} - \ket{1} \right) \left(\sqrt{r} \bra{0} - \bra{1} \right), \\
\widetilde{N}_i &=& (1-r) \ket{0} \bra{0}, \\
\widetilde{N}_j &=& \ket{2} \bra{2}.
\end{eqnarray*}
Note that by symmetry, we only need explicitly consider one value of $i$ and $j$. For $i=6$, $j=8$, we obtain 
\begin{eqnarray*}
\widetilde{\sigma} &=& \frac{1}{2} (1+ \sqrt{r} )^2 \ket{0} \bra{0} + (1-r) \ket{1+2} \bra{1+2} \\
&&+ \ket{0+1} \bra{0+1} \\
& \leq & (1+r) \ket{0} \bra{0} + (1-r) \ket{1+2} \bra{1+2} \\
&& + \ket{0+1} \bra{0+1}, \\
& \leq & (1+r) \ket{0} \bra{0} + (1-r) \ket{1} \bra{1}+ (1-r) \ket{2} \bra{2}\\
&& + \ket{0+1} \bra{0+1},
\end{eqnarray*}
where in the second line we have used the inequality $(a+b)^2 \leq 2 (a^2 + b^2)$, and in the third line we have added the positive operator $(1-r) \ket{1-2} \bra{1-2}$. Thus in matrix notation, in the $\{ \ket{0}, \ket{1}, \ket{2} \}$ basis, we obtain
\begin{eqnarray*}
&& \Gamma^A - \frac{1}{9} \widetilde{\sigma} \geq \\
&& \left( \begin{array}{ccc} p - \frac{1}{9}(1+r) - \frac{1}{18} & -\frac{1}{18} & 0 \\ -\frac{1}{18} & q - \frac{1}{9}(1-r) - \frac{1}{18} & 0 \\ 0 & 0 & p-\frac{1}{9}(1-r) \end{array} \right).
\end{eqnarray*}
The diagonal elements are strictly positive, while $\left( p - \frac{1}{9}(1+r) - \frac{1}{18})(q-\frac{1}{9}(1-r)-\frac{1}{18} \right) - \left(\frac{1}{18}\right)^2 \geq 0$ according to the discussion in Case 1 above. Thus the operator inequality holds.

{\bf Case 3}: $\tan^2 \alpha + \tan^2 \beta < 1$:

All measurements in which $\{ \widetilde{N}_1, \widetilde{N}_2, \widetilde{N}_3, \widetilde{N}_4, \widetilde{N}_5 \}$ are the only non-zero elements may be parametrised in the following way, where $r = \tan^2 \alpha$, $0 \leq r < 1$, $s= \tan^2 \beta$, $0 \leq s <1$:
\begin{eqnarray*}
\widetilde{N}_1 &=& (1-r-s) \ket{1} \bra{1}, \\
\widetilde{N}_2 &=& \frac{1}{2} \left(\ket{0} + \sqrt{r} \ket{1} \right) \left(\bra{0} + \sqrt{r} \bra{1} \right), \\
\widetilde{N}_3 &=& \frac{1}{2} \left(\ket{0} - \sqrt{r} \ket{1} \right) \left(\bra{0} - \sqrt{r} \bra{1} \right), \\
\widetilde{N}_4 &=& \frac{1}{2} \left(\ket{2} + \sqrt{s} \ket{1} \right) \left(\bra{2} + \sqrt{s} \bra{1} \right), \\
\widetilde{N}_5 &=& \frac{1}{2} \left(\ket{2} - \sqrt{s} \ket{1} \right) \left(\bra{2} - \sqrt{s} \bra{1} \right).
\end{eqnarray*}
From eqn (\ref{sigmatilde}) we obtain 
\begin{eqnarray*}
\widetilde{\sigma} &=& (1-r-s) \ket{1} \bra{1} + \frac{1}{2} (1+ \sqrt{r} )^2 \ket{0} \bra{0} \\
&&+ \frac{1}{2} (1+ \sqrt{s} )^2 \ket{2} \bra{2} \\
& \leq & (1+r) \ket{0} \bra{0} + (1-r-s) \ket{1} \bra{1} + (1+s) \ket{2} \bra{2}.
\end{eqnarray*}
where we have used again the inequality $(a+b)^2 \leq 2 (a^2 + b^2)$. This is diagonal in the $\{ \ket{0}, \ket{1}, \ket{2} \}$ basis, and it is easy to see that $\Gamma^A - \frac{1}{9} \widetilde{\sigma} \geq 0$ holds.

{\bf Case 4}: $\tan^2 \alpha + \tan^2 \beta > 1$:

All measurements in which $\{ \widetilde{N}_2, \widetilde{N}_3, \widetilde{N}_4, \widetilde{N}_5, \widetilde{N}_i \}$, $i=6,7$, are the only non-zero elements may be parametrised in the following way, where $r = \frac{1}{\tan^2 \alpha}$, $0 \leq r < 1$, $s= \tan^2 \beta$, $0 \leq s <1$:
\begin{eqnarray*}
\widetilde{N}_2 &=& \frac{1-s}{2} \left(\sqrt{r} \ket{0} + \ket{1} \right) \left(\sqrt{r} \bra{0} + \bra{1} \right), \\
\widetilde{N}_3 &=& \frac{1-s}{2} \left(\sqrt{r} \ket{0} - \ket{1} \right) \left(\sqrt{r} \bra{0} - \bra{1} \right), \\
\widetilde{N}_4 &=& \frac{1}{2} \left(\ket{2} + \sqrt{s} \ket{1} \right) \left(\bra{2} + \sqrt{s} \bra{1} \right), \\
\widetilde{N}_5 &=& \frac{1}{2} \left(\ket{2} - \sqrt{s} \ket{1} \right) \left(\bra{2} - \sqrt{s} \bra{1} \right), \\
\widetilde{N}_i &=& (1-r(1-s)) \ket{0} \bra{0},
\end{eqnarray*}
From eqn (\ref{sigmatilde}) we obtain 
\begin{eqnarray*}
\widetilde{\sigma} &=& \frac{1-s}{2} (1+ \sqrt{r} )^2 \ket{0} \bra{0} + \frac{1}{2} (1+ \sqrt{s} )^2 \ket{2} \bra{2} \\
&& + (1-r(1-s)) \ket{1+2} \bra{1+2} \\
& \leq & (1-s)(1+r) \ket{0} \bra{0}+ (1+s) \ket{2} \bra{2} \\
&& + (1-r(1-s)) \ket{1+2} \bra{1+2}, \\
& < & (1-s)(1+r) \ket{0} \bra{0} \\
&& 2 \ket{2} \bra{2} + \ket{1+2} \bra{1+2},
\end{eqnarray*}
where we have used again the inequality $(a+b)^2 \leq 2 (a^2 + b^2)$, and in the last line use $s<1$. Recall that $\Gamma^A$ is diagonal in the $\{ \ket{0}, \ket{1}, \ket{2} \}$ basis, so it suffices to check
$$
\Gamma^A - \frac{1}{9} (1-s)(1+r) \ket{0} \bra{0} \geq 0,
$$
which clearly holds as $\frac{1}{9}(1-s)(1+r) < \frac{2}{9} < p$, and
$$
\Gamma^A - \frac{2}{9} \ket{2} \bra{2} + \frac{1}{9} \ket{1+2} \bra{1+2} \geq 0,
$$
which we recognise as $\Gamma^A - \frac{1}{3} \rho_7 \geq 0$, and therefore holds by construction.

\end{document}